\pdfoutput=1
\documentclass[11pt,a4paper]{article}

\usepackage{jcappub2}
\usepackage[latin1]{inputenc}
\usepackage{bbm}
\usepackage{graphicx}
\usepackage{epsfig}
\usepackage{subfigure}
\usepackage{color}
\usepackage{slashed}

\newcommand{\be}{\begin{equation}} 
\newcommand{\ee}{\end{equation}}
\newcommand{\bea}{\begin{equation}\begin{aligned}} 
\newcommand{\eea}{\end{aligned}\end{equation}}

\newcommand{\bmp}{\noindent\begin{minipage}{16cm}}
\newcommand{\emp}{\end{minipage}\vskip 7mm} 
\def\lsim{\mathrel{\raise.3ex\hbox{$<$\kern-.75em\lower1ex\hbox{$\sim$}}}}
\def\gsim{\mathrel{\raise.3ex\hbox{$>$\kern-.75em\lower1ex\hbox{$\sim$}}}}

\newcommand{\intron}[1]{}

\newcommand{\GeV}{\,{\mathrm{GeV}}}
\renewcommand{\i}{\mathrm{i}}

\def\muHS{\mu_{\hspace{-0.2mm}\vphantom{\frac{1}{2}}\mathit{HS}}} 
\def\lHS{\lambda_{\hspace{-0.2mm}\mathit{HS}}} 
\def\Tc{T_{\mathrm{c}}}

\subheader{\small{preprint: HIP-2014-12/TH}}

\title{Strong phase transition, dark matter and vacuum stability from simple hidden sectors}
\author[a,c]{Tommi Alanne,}
\author[b,c]{Kimmo Tuominen} 
\author[a,c]{and Ville Vaskonen}

\affiliation[a]{Department of Physics, University of Jyv\"askyl\"a, \\
                      P.O.Box 35 (YFL), FI-40014 University of Jyv\"askyl\"a, Finland}
\affiliation[b]{Department of Physics, University of Helsinki, \\
                      P.O.Box 64, FI-00014 University of Helsinki, Finland}                 
\affiliation[c]{Helsinki Institute of Physics, \\
                      P.O.Box 64, FI-00014 University of Helsinki, Finland}
\emailAdd{tommi.alanne@jyu.fi}
\emailAdd{kimmo.i.tuominen@helsinki.fi}
\emailAdd{ville.vaskonen@jyu.fi}

\abstract{Motivated by the possibility to explain dark matter abundance and strong electroweak phase transition, we consider simple extensions of the Standard Model containing singlet fields coupled with the Standard Model via a scalar portal. Concretely, we consider a basic portal model consisting of a singlet scalar with $Z_2$ symmetry and a model containing a singlet fermion connected with the Standard Model fields via a singlet scalar portal. We perform a Monte Carlo analysis of the parameter space of each model, and we find that in both cases the dark matter abundance can be produced either via freeze-out or freeze-in mechanisms, but only in the latter model one can obtain also a strong electroweak phase transition required by the successful electroweak baryogenesis. We impose the direct search limits 
and consider systematically the possibility that the model produces only a subdominant portion of the dark matter abundance. We also study the renormalization group evolution of the couplings of the model to determine if the scalar sector of the model remains stable and perturbative up to high scales. With explicit examples of benchmark values of the couplings at weak scale, we show that this is possible. Models of this type are further motivated by the possibility that the excursions of the Higgs field at the end of inflation are large and could directly probe the instability region of the Standard Model.}


\begin{document}
\maketitle
\section{Introduction}
Extensions of the Standard Model (SM) of elementary particle interactions have been put under severe
tests \cite{Carmi:2012yp, Espinosa:2012ir, Giardino:2012ww, Alanne:2013dra} after the discovery of the Higgs boson with mass $m_h=126$ GeV at the ATLAS and CMS experiments in the CERN Large Hadron Collider \cite{Aad:2012tfa,Chatrchyan:2012ufa}. The requirement of a light Higgs scalar boson and no other obviously accessible states at the energy scales probed so far present a challenge for traditional model paradigms like supersymmetry and technicolor which predict an extended spectrum beyond the SM. On the other hand, the cosmological observations on the dark matter abundance and matter-antimatter asymmetry clearly require, in the elementary particle physics context, the existence of new degrees of freedom not present in the SM. 

One possible framework to address these aspects is to take the SM according to the current collider data and extend it with singlet fields communicating with the SM fields only through the scalar or vector portals. 
The singlet sector can consist of a single scalar \cite{McDonald:1993ex, Burgess:2000yq}, 
more complex scalar multiplets  \cite{LopezHonorez:2006gr,Alvares:2012qv}, fermions 
\cite{LopezHonorez:2012kv, Fairbairn:2013uta,Alves:2013tqa} or vectors \cite{Hambye:2008bq, Davoudiasl:2013jma}. 
The resulting spectrum can contain particles stable over the 
timescales of the age of the universe and contribute to the observed abundance of the dark matter. 
Moreover, the extended scalar potential can modify the properties of the phase transitions in the early universe with respect to the results obtained in the SM. A strong first-order electroweak transition is a prerequisite for successful electroweak baryogenesis \cite{Kuzmin:1985mm}, and it is well known that the electroweak phase transition in the SM is not of first order but a smooth crossover \cite{Kajantie:1996mn,Rummukainen:1998as}. If the electroweak sector of the SM were fully perturbative, 
a first-order phase transition would arise from a cubic term generated in the Higgs effective one-loop potential by the thermal effects of fields coupled to the Higgs. However, addition of a singlet scalar can sufficiently modify the picture already by tree level effects due to the presence of $T$-independent dimensional parameters appearing in the scalar potential and lead to a strong first-order transition \cite{Profumo:2007wc, Espinosa:2011ax}. 
Consequently, the ratio $v(\Tc)/\Tc$ which controls the sphaleron erasure of the baryon asymmetry can be large and lead to succesful electroweak baryogenesis.

The phenomenologically interesting scenario would, thus, be the one where the strong electroweak phase transition is accompanied by an explanation of the dark matter relic density by a weakly coupled massive particle  (WIMP). As a concrete model paradigm in this paper, we consider simple scalar portals between the hidden sector and the visible one. In addition to the Standard Model Higgs field, the scalar sector contains a real singlet scalar $S$. 
 
As a simple limiting case, we consider a model where the singlet sector is solely constituted by $S$ with a discrete $Z_2$ symmetry \cite{McDonald:1993ex,McDonald:2001vt,Cline:2012hg,Cline:2013gha}. 
Then $S$ can also act as a dark matter candidate provided that the global minimum of the potential at zero temperature does not spontaneously break this $Z_2$ symmetry. However, in this model it is not possible to simultaneously explain the strong electroweak phase transition and the observed dark matter relic abundance. Therefore, as a second example we consider a model where the scalar $S$ is not assumed to have any discrete symmetries, but the singlet sector also contains a Dirac fermion \cite{LopezHonorez:2012kv, Fairbairn:2013uta,Li:2014wia}. We find that in this model it is possible to realize simultaneously strong electroweak phase transition and the observed dark matter relic abundance. 
 
To establish our results, we perform a Monte Carlo analysis of the parameter space to search for viable models. We impose the constraints from LHC data and precision electroweak measurements, and from the 
dark matter direct searches \cite{Aprile:2011hi, Akerib:2013tjd}. 
To improve the earlier work on models of this type \cite{Cline:2012hg,Fairbairn:2013uta}, 
we also require that the couplings remain perturbative up to scales of ${\cal O}($TeV$)$. 
As a special case, we consider the possibility that the model could remain perturbative up to the Planck scale. 
This happens only on very specific couplings and allows to single out specific benchmark models where 
the dark matter constraints can be satisfied and a strong electroweak transition induced. 
Furthermore, since it is possible that multiple components contribute to the observed dark matter abundance
\cite{Duda:2001ae,Profumo:2009tb,Aoki:2012ub}, we consider systematically the cases of subdominant dark matter and both freeze-out and freeze-in mechanisms to generate the dark matter abundance. In particular, we show that if the dark matter candidate is a singlet fermion, then its abundance can be generated either 
via the freeze-out or the freeze-in mechanism, and in both cases the model will also lead to a strong electroweak phase transition.

The paper is organized as follows: In section \ref{model}, we introduce the model and the basic constraints. Then in section \ref{results}, we discuss the numerical results for the model with only a $Z_2$-symmetric real singlet scalar and for the model with a real singlet scalar and a singlet Dirac fermion. In section \ref{checkout}, we present our conclusions.

\section{Model and constraints}
\label{model}
We consider an extended scalar sector described by the potential
\bea
V(H,S)=&\mu_H^2H^\dagger H+\lambda_H(H^\dagger H)^2+\frac{1}{2}\mu_S^2 S^2+\frac{\mu_3}{3} S^3+\frac{\lambda_S}{4}S^4\\
&+\muHS (H^\dagger H)S+\frac{\lHS}{2}(H^\dagger H)S^2,
\label{scalarpotII}
\eea
which provides typical scalar portals between the hidden singlet sector and the Standard Model Higgs.
In Eq. (\ref{scalarpotII}), the field $H$ is the usual Standard Model Higgs doublet and $S$ is a real singlet scalar. The Higgs doublet is written in terms of the components as
\be
H=\begin{pmatrix} \phi^+\\ \frac{1}{\sqrt{2}}(v+\phi_{\mathrm{r}}^0+\mathrm{i}\phi_{\mathrm{i}}^0)\end{pmatrix},
\ee
where the superscript refers to the electroweak charge of the component and subscripts denote the real and imaginary parts. Note that the most general potential would have a linear term, $\mu_1^3S$, but this can always be removed by shifting the field $S$. 
The stability of the potential requires that 
\be
\lambda_H>0, \quad \lambda_S>0, \quad \lHS>-2\sqrt{\lambda_H \lambda_S}.
\label{vacstab}
\ee

Depending on the values of the parameters, the vacuum structure can be rich \cite{Espinosa:2011ax}. We require that the extremum which leads to correct pattern of the electroweak symmetry breaking is the global minimum at zero temperature.

To study the consequences for the electroweak phase transition, we need to include the finite-temperature corrections. Since there are multiple vacua induced already at the tree level, it is sufficient to consider the finite-temperature corrections to the leading terms. In other words, we consider temperature dependent coefficients
\be
\mu_1(T)^3=c_1 T^2,\quad \mu_S(T)^2=\mu_S^2+c_S T^2,\quad \mu_H(T)^2=\mu_H^2+c_H T^2,
\label{thermalcorr}
\ee
where
\bea
c_1 &= \frac{1}{12}(\mu_3 + 2\muHS), \\
c_S &= \frac{1}{12}(2\lHS+3\lambda_S), \\
c_H &= \frac{1}{48}(9 g^2+3g^{\prime 2}+12 y_t^2+24\lambda_H+2\lHS),
\label{thermalc}
\eea
in Eq. (\ref{scalarpotII}).  We have assumed that the term linear in $S$, i.e. the one with coefficient $\mu_1$ in the potential, has been shifted away at tree level at zero temperature and therefore arises only for $T\neq 0$.
Starting from $T=0$, we monitor the evolutions of the electroweak symmetric and electroweak broken minima. If the symmetric minimum at some temperature $\Tc$ gets deeper than the asymmetric minimum, the phase transition is possible. For these cases we determine the critical temperature, $\Tc$, and $v(\Tc)/\Tc$ to identify the parameter space domains where a strong electroweak transition can be realized.

For the dark matter candidate, there are several possibilities which can be built around this scalar sector. We will consider two. The simplest, and already much studied, case is to assume the singlet scalar to have a discrete symmetry, like $Z_2$, which renders it stable. This symmetry must then be imposed on the general potential of Eq. (\ref{scalarpotII}).  The second alternative that we consider is to assume, in addition to the scalar, the existence of a singlet fermion, which then due to a conserved fermion number becomes a stable dark matter candidate. Then no symmetry restrictions need to be imposed on the scalar potential, leaving a richer possibility of patterns for the phase transition between the electroweak symmetric and broken vacua. The details and numerical result of these two model examples will be exposed more thoroughly in the next section. Here, we now discuss the general formulation of the computation of the dark matter abundance and direct detection limits.

To compute the relic abundance of dark matter, $\Omega_{\rm{DM}}$, we apply the standard freeze-out formalism.\footnote{Of course, dark matter abundance need not be due to a thermal relic. We will briefly discuss the possibility of producing the abundance via out-of-equilibrium freeze-in scenario in the models we consider when presenting the numerical results in Sec. \ref{results}.} The number density $n$ of the thermal relic can be computed from the Lee-Weinberg equation \cite{Lee:1977ua},
 \be
 \frac{\partial f(x)}{\partial x}=\frac{\langle v\sigma\rangle m^3 x^2}{H}(f^2(x)-f_{\rm{eq}}^2(x)),
 \ee
 written in terms of scaled variables $f(x)=n(x)/s_E$ and $x=s_E^{1/3}/m$. Here $s_E$ is the entropy density at temperature $T$, $m$ is the mass of the dark matter candidate and $H$ is the Hubble parameter. For the averaged cross sections, we use the integral expression \cite{Gondolo:1990dk}
 \be
 \langle v\sigma\rangle=\frac{1}{8m^4 T K_2^2(m/T)}\int_{4 m^2}^\infty ds\sqrt{s} (s-4 m^2)K_1(\sqrt{s}/T)\sigma_{\rm{tot}}(s),
 \ee
 where $K_i(y)$ are the modified Bessel functions of the second kind and $s$ is the usual Mandelstam invariant. Given the cross sections, we can solve the Lee-Weinberg equation for $f(0)$ which gives the present ratio of the number density of the dark matter candidate to its entropy density. The fractional density parameter, $\Omega_{\rm{DM}}$, can be computed from
 \be
 \Omega_{\rm{DM}}h^2\simeq 2.76\cdot 10^8\, \left(\frac{m}{\GeV}\right) f(0).
 \ee
Since there is no reason to expect that all of the dark matter abundance originates from a single source, we define the fraction
\be
f_{\rm{rel}}=\Omega_{\rm{DM}} h^2/(\Omega h^2)_{\mathrm{c}},
\label{freldef}
\ee
where $(\Omega h^2)_{\mathrm{c}}=0.12$ from Planck \cite{Ade:2013zuv}.

Also, there are stringent constraints for this type of models arising both from the LHC data and direct searches for dark matter. If the singlet scalar is light enough, $2m_S\le m_h$, then an obvious constraint on new light degrees of freedom from LHC data is the invisible width of the Higgs boson. According to present LHC data \cite{ATLAS-CONF-2014-009, Chatrchyan:2013zna, Chatrchyan:2014nva, CMS:ril, Chatrchyan:2013iaa, Chatrchyan:2013mxa}, the $2\sigma$ limit for the branching fraction to invisible channels is ${\rm{Br}}_{\rm{inv}}\le 0.28$. On the other hand, after the symmetry breaking the scalar mass eigenstates are 
\be
h^0 = \phi_{\mathrm{r}}^0 \cos\beta + S \sin\beta, \quad H^0 = -\phi_{\mathrm{r}}^0 \sin\beta + S \cos\beta,
\label{scalarmixing}
\ee
and the mixing of the Higgs and the singlet scalar affects the couplings of the scalar to fermions and gauge bosons. 
These modifications can be constrained by the current collider data by fitting the mixing $\cos\beta$ to the signal strength results.
Moreover, compatibility with the electroweak precision measurements using the $S$ and $T$ parameters \cite{Peskin:1990zt} needs to be checked.  Finally, considering the direct searches for dark matter, the scalar portal interactions in Eq. (\ref{scalarpotII}) contribute to the spin-independent scattering cross sections on nuclei for which the LUX experiment \cite{Akerib:2013tjd} provides currently the most stringent constraints.

In the cases we will consider, the WIMP couples to the nucleus via the scalars $h^0$ and $H^0$ with strength depending on the mixing patter of the scalars and whether the WIMP is a scalar or a fermion. In both cases the Higgs-nucleon coupling is $f_N m_N/v$, where $m_N=0.946$ GeV, and we neglect the small differences between neutrons and protons.
The effective Higgs-nucleon coupling, 
\be
f_N\equiv \frac{1}{m_N}\sum_q\langle N|m_q\bar{q}q| N\rangle,
\ee
describes the normalized total quark scalar current within the nucleon. The quark currents of the nucleon have been a subject of an intensive lattice research supplemented by efforts applying chiral perturbation theory methods and pion nucleon scattering. Consequently, $f_N$ is fairly well determined currently. Following \cite{Cline:2013gha} we use $f_N=0.345\pm 0.016$, where the uncertainty in $f_N$ induces at most 20\% error in the spin-independent direct detection limits.

The spin-independent cross section for a WIMP scattering on nuclei is computed by considering the $t$-channel exchange of $h^0$ and $H^0$ in the limit $t\rightarrow 0$. The matrix element for this process generally contains the factors
\be
\left(\frac{\tilde{g}_h\cos\beta}{m_h^2} - \frac{\tilde{g}_H\sin\beta}{m_H^2} \right)f_N\frac{m_N}{v},
\label{DMeffcoupl}
\ee
where the coupling between the scalars $h^0$ and $H^0$  and the WIMP candidate is denoted 
by $\tilde{g}_h$ and $\tilde{g}_H$, respectively. In the explicit models we consider, 
we determine $\tilde{g}_h$ and $\tilde{g}_H$ and use the above formula when evaluating the spin-independent scattering cross section per nucleon, $\sigma_{\rm{SI}}^0$. 
Since we consider the possibility that our WIMP candidate forms only a fraction of the total dark matter abundance, we need to take this into account when comparing with the direct searches. The direct search constraints on $\sigma^0_{\rm{SI}}$ are given by the experimental collaborations under the assumption that $f_{\rm{rel}}=1$. To apply the constraints under the assumption of subdominant WIMPs, we define an effective cross section
\be
\sigma_{\rm{SI}}^{\rm{eff}}=f_{\rm{rel}}\sigma^0_{\rm{SI}}.
\ee

With these preliminaries we now turn to the numerical results of two explicit models based on the scalar sector described by the potential (\ref{scalarpotII}).

\section{Numerical results}
\label{results}
We will now discuss two different concrete cases: First, assuming $Z_2$ symmetry for the singlet scalar simplifies the potential and allows the singlet scalar to act as dark matter. However, it is impossible to saturate even a modest fraction of the observed dark matter abundance and simultaneously induce a strong first-order electroweak phase transition. Second, we release the requirement of $Z_2$ symmetry on the singlet scalar but assume the existence of a singlet fermion. We find that in this case the scalar can induce a strong first-order electroweak transition, and the fermion can act as a dark matter candidate.

\subsection{Scalar dark matter: $Z_2$-symmetric case}
A simple benchmark model, which has been considered extensively in literature earlier \cite{McDonald:1993ex,McDonald:2001vt,Cline:2012hg,Cline:2013gha}, is provided by imposing a $Z_2$ symmetry on the potential in Eq. (\ref{scalarpotII}).

Consider the Lagrangian 
\be
{\cal L}={\cal L}_{\rm{kin}}+{\cal L}_{\rm{Yuk}}-V(H,S)\vert_{Z_2},
\ee
where ${\cal L}_{\rm{kin}}$ contains the kinetic terms for the scalars, appropriately gauged under the SM charges, and for all the SM gauge and matter fields. The term ${\cal L}_{\rm{Yuk}}$ contains the usual 
Yukawa couplings between the SM matter fields and the Higgs field $H$. The potential $V(H,S)\vert_{Z_2}$ is obtained from Eq. (\ref{scalarpotII}) by setting $\muHS=\mu_3=0$. 
Due to the $Z_2$ symmetry, the singlet scalar in this model can act as a dark matter candidate.
Depending on the model parameters, either the neutral component of $H$ or the singlet can have a nonzero vacuum expectation value (vev). Since we assume the singlet to be a dark matter candidate, the $T=0$ vacuum has to be $Z_2$ symmetric to ensure the stability of the dark matter candidate. This means that at zero temperature the vev, $w$, of $S$ must be zero, i.e. the global minimum of the potential must be at  $(v,w)=(246$ GeV, $0)$. This requires, in addition to the bounds from vacuum stability in Eq. (\ref{vacstab}), that
\be
\mu_S^2 > -v^2\sqrt{\lambda_H \lambda_S}.
\ee
In the notation of Eq. (\ref{scalarmixing}), the mass eigenstates are directly $h^0=\phi_r$ and $H^0=S$, i.e. $\cos\beta=1$. We trade the mass parameter $\mu_S^2$ with the physical mass of the singlet, $m_S^2=\mu_S^2+v^2\lHS/2$. Then the above constraint and the requirement of the stability can be combined into a bound
\be
-2 \sqrt{\lambda_H\lambda_S}< \lHS<\frac{2m_S^2}{v^2} + 2 \sqrt{\lambda_H \lambda_S}.
\ee

The known values of the electroweak scale, $v=246$ GeV, and the Higgs mass, $m_h=126$ GeV, fix the model parameters
$\lambda_H=0.131$ and 
$\mu_H^2=-v^2\lambda_H$.
The remaining parameters, $m_S^2$, $\lambda_S$ and $\lHS$ are free but subject to the constraints listed above. We scan the parameter space by performing a simple Monte Carlo analysis generating a random distribution of points with
\be
0 < \lambda_S\le \pi,\quad 5\GeV\le m_S\le 650\,{\textrm{ GeV}}.
\ee

As a further constraint, we require perturbativity of the couplings reasonably far out in energy. 
The renormalization group (RG) equations are solved at one loop, and points which remain perturbative up to scales $\mu\sim 10$ TeV are accepted. The RG equations for the gauge couplings and the Yukawa coupling of the top quark are as in the Standard Model. The $\beta$~functions of the couplings are defined as
\begin{equation}
    \beta_{g}=\frac{\mathrm{d}g}{\mathrm{d}\ln\mu},
\end{equation}
and in the $Z_2$-symmetric case they read
\begin{align}
    \begin{split}
	16\pi^2 \beta_{\lambda_H}=& 24\lambda_H^2+\frac{1}{2}\lHS^2-3\left(3g^2
	    +g^{\prime 2}-4y_t^2\right)\lambda_H\\
	&+\frac{3}{8}\left(3g^4+2g^2g^{\prime 2}+g^{\prime 4}\right)-6y_t^4, 
    \end{split} \\
    \begin{split}
	16\pi^2\beta_{\lHS}=& 4\lHS^2+\left(12\lambda_H+6\lambda_S\right)\lHS\\
	&-3\left(\frac{3}{2}g^2+\frac{1}{2}g^{\prime 2}-2y_t^2\right)\lHS,
    \end{split} \\
    \begin{split}
	16\pi^2\beta_{\lambda_{S}}=&18\lambda_S^2+2\lHS^2.
    \end{split}
\end{align}

To determine how much of the dark matter relic abundance the singlet scalar $S$ can provide for, we 
carry out the standard freeze-out calculation. There are three annihilation channels, $SS\rightarrow h^0h^0, VV$ and $\bar{f}f$, where $V$ denotes the electroweak gauge bosons, $V=W,\,Z$. The annihilation cross sections to these three distinct final states are
\bea
\sigma_{hh} &= \frac{v_h}{32\pi s v_S}\left|\lHS+\frac{3m_h^2\lHS}{s-m_h^2+\i m_h\Gamma_h}-\frac{4v^2\lHS^2}{s-2m_h^2}\right|^2, \\
\sigma_{VV} &= \frac{v_V}{4\pi sv_S}\frac{M^4_V\lHS^2}{|s-m_h^2+\i m_h\Gamma_h|^2}\left(3+\frac{s(s-4M_V^2)}{4M_V^4}\right)\delta_V, \\
\sigma_{ff} &= \frac{v_f X_f}{16\pi sv_S}\frac{m_f^2(s-4m_f^2)\lHS^2}{|s-m_h^2+\i m_h\Gamma_h|^2}.
\eea
Here $v_X=\sqrt{1-4m_X^2/s}$, $\delta_{W,Z}=1,1/2$ and for QCD colour singlet fermions $X_f=1$ while for quarks
\be
X_f=3\left(1+\left(\frac{3}{2}\ln\frac{m_q^2}{s}+\frac{9}{4}\right)\frac{4\alpha_s}{3\pi}\right),
\ee
where $\alpha_s=0.12$ is the strong coupling constant.\footnote{For the importance of this QCD correction, see Appendix A of \cite{Cline:2013gha}.}

In the range $m_h/2<m_S<m_h$, we factorize the annihilation to fermion and gauge boson final states to the $SS\to h^0$ fusion part times the virtual Higgs decay using the full width of the Higgs \cite{Dittmaier:2011ti}, which also takes the 4-body final states into account. 

To see if this model can provide for the strong electroweak transition, $v(\Tc)/\Tc>1$, we consider the finite-temperature corrections to the quadratic terms; the linear term in $S$ does not arise in this $Z_2$-symmetric case.
At very high temperatures, the potential has a unique minimum at $(S,H)=(0,0)$. As the temperature decreases, the potential generates two minima: one at nonzero $S$ and the other at nonzero value of the neutral component of $H$. At some intermediate temperature these two become degenerate, and the possibility for a strong electroweak phase transition arises.

To take into account the constraints from LHC, we consider the decay width of the Higgs to two singlets, 
$h^0\rightarrow SS$,
\be
\Gamma_{h^0\rightarrow SS}=\frac{\lHS^2 v^2}{32\pi m_h} \sqrt{1-\frac{4m_S^2}{m_h^2}}.
\label{higgsdecaywidth}
\ee
The Higgs total decay width to the visible Standard Model channels is 
$\Gamma_h=4.07$ MeV for $m_h=126$ GeV \cite{Dittmaier:2011ti}, and this implies a bound for the Higgs portal coupling. 

\begin{figure}
\includegraphics[width=0.5\textwidth]{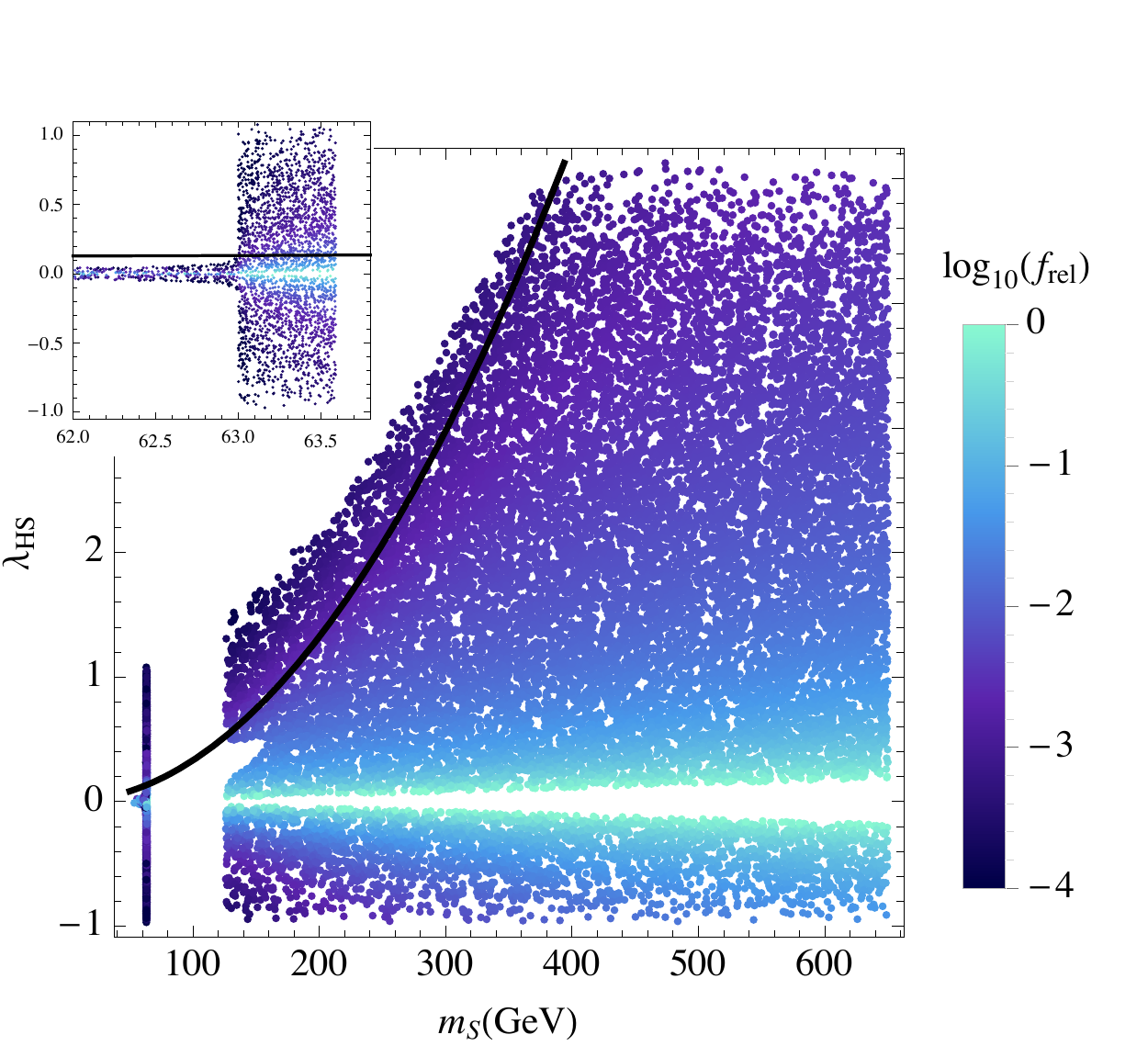}
\includegraphics[width=0.5\textwidth]{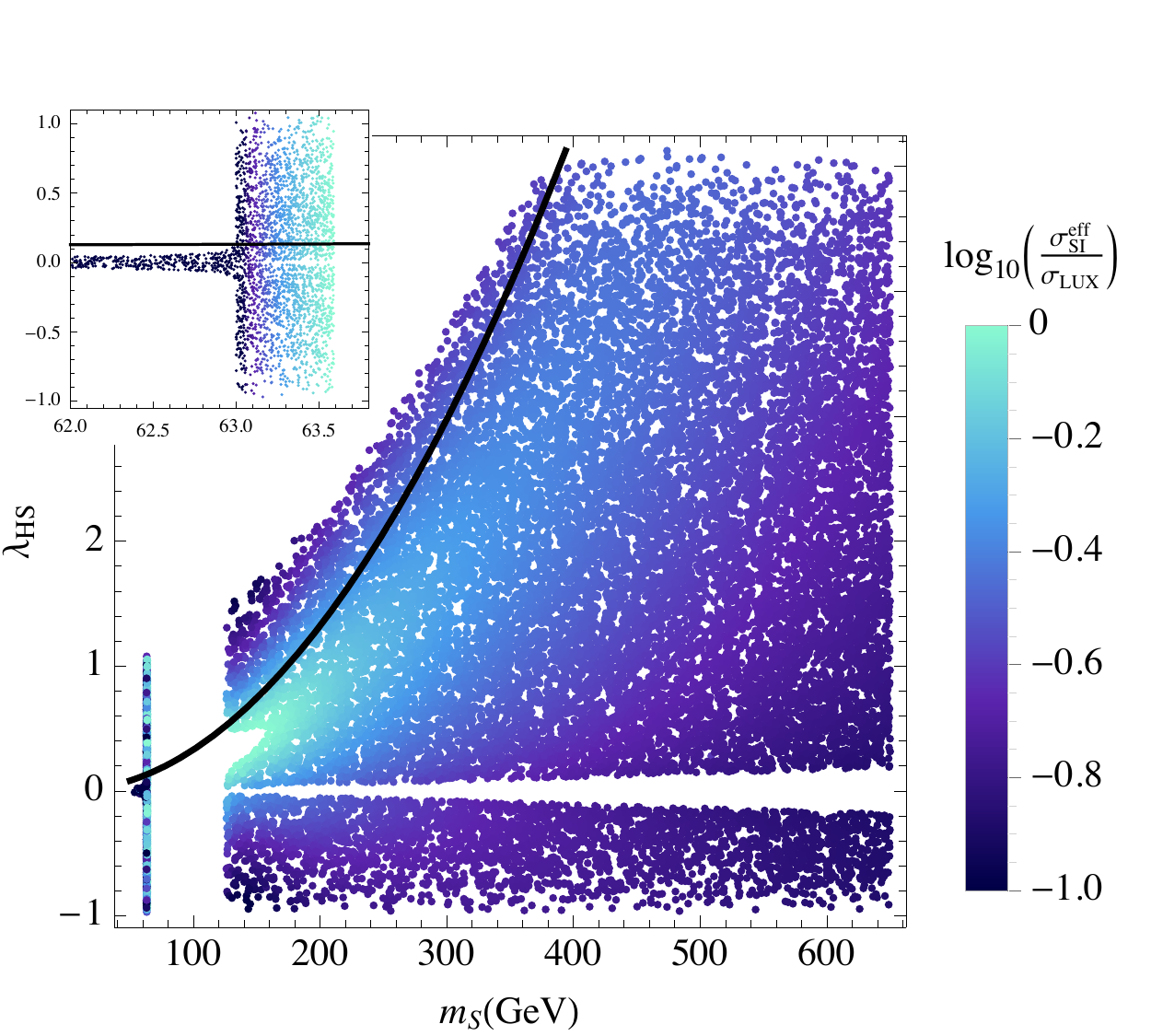}
\caption{Left panel: the color coding shows the value of $f_{\rm{rel}}$ in the $(m_S,\lHS)$ plane. Above the solid black line $\mu_S^2<0$, and modification to the electroweak transition are possible. The constraint from the perturbativity of couplings, LHC invisible Higgs width and direct dark matter searches have been imposed as explained in the text. Right panel: results from dark matter direct searches. The color coding shows how the parameter space is constrained further if the sensitivity of the direct search experiments rises. For the black points $\sigma_{\rm{SI}}^{\rm{eff}}/\sigma_{\rm{LUX}} \le 0.1$.}
\label{basic}
\end{figure}

The basic result of the model is shown in the left panel of Fig. \ref{basic}. The color coding shows the relative dark matter relic abundance. Above the solid black curve, $\lHS=2 m_S^2/v^2$,  $\mu_S^2$ is negative and modifications to the electroweak transition are possible. The figure illustrates that the parameter space of the model where (a reasonable fraction of) the relic density can be explained does not overlap with the parameter space where a strong first-order phase transition may arise.

The points in the left panel of Fig. \ref{basic} have also been constrained to be compatible with the direct search results from the LUX experiment \cite{Akerib:2013tjd}, which provides currently the best constraints for this type of model where only spin-independent cross section between the WIMP and ordinary matter arises. In the model we consider here, the effective couplings $\tilde{g}_h$ and $\tilde{g}_H$ in Eq. (\ref{DMeffcoupl}) are
\be
\tilde{g}_h=\frac{\lHS}{2}v, \; \tilde{g}_H = 0.
\ee
Using these, we evaluate the cross section for elastic WIMP nucleon scattering and obtain
\be
\sigma_{\rm{SI}}^0 = \frac{1}{4\pi}\frac{\lHS^2 \mu_N^2 f_N^2 m_N^2}{m_h^4 m_S^2},
\ee
where $\mu_N^2$ is the reduced mass of the WIMP-nucleon pair. To illustrate the distance of our results from the LUX bound, we show in the right panel of Fig. \ref{basic} the same points as in the left panel, with the color coding now showing the spin-independent cross section in relative to the LUX constraint.

Of course the dark matter candidate need not be produced by the freeze-out mechanism as we have assumed so far. Since it is impossible to explain both the strong electroweak phase transition and the dark matter abundance with the scalar sector considered here, let us leave the phase transition for a while and focus only on dark matter. Given the fact that no trace of WIMPs has been observed in the direct searches, we consider briefly the possibility of freezing in the relic abundance \cite{Hall:2009bx}. Assuming that the scalar is light, it can be produced from the thermal bath of Higgs bosons. The number density of $S$ is described by the Boltzmann equation formally similar to the one governing the density in the freeze-out case. 
Analogously to the well known approximate result in the freeze-out case, the dark matter abundance in this case is given by
\be
\Omega_S h^2\simeq \frac{1.09\cdot 10^{27}}{g_s\sqrt{g_\rho}}\frac{m_S\Gamma_{h^0\rightarrow SS}}{m_h^2},
\ee
where $g_{s,\rho}$ denotes the effective degrees of freedom for the entropy and energy density, respectively. The essential feature of this mechanism is that the coupling required for the production of sufficient relic abundance is superweak; one obtains a parametric estimate
\be
\lHS\simeq 2\times10^{-11} \left(\frac{\Omega_Sh^2}{0.12}\right)^{1/2}\left(\frac{{\rm{GeV}}}{m_S}\right)^{1/2}.
\label{freezeinlvsm}
\ee

\begin{figure}
\begin{center}
\includegraphics[width=0.6\textwidth]{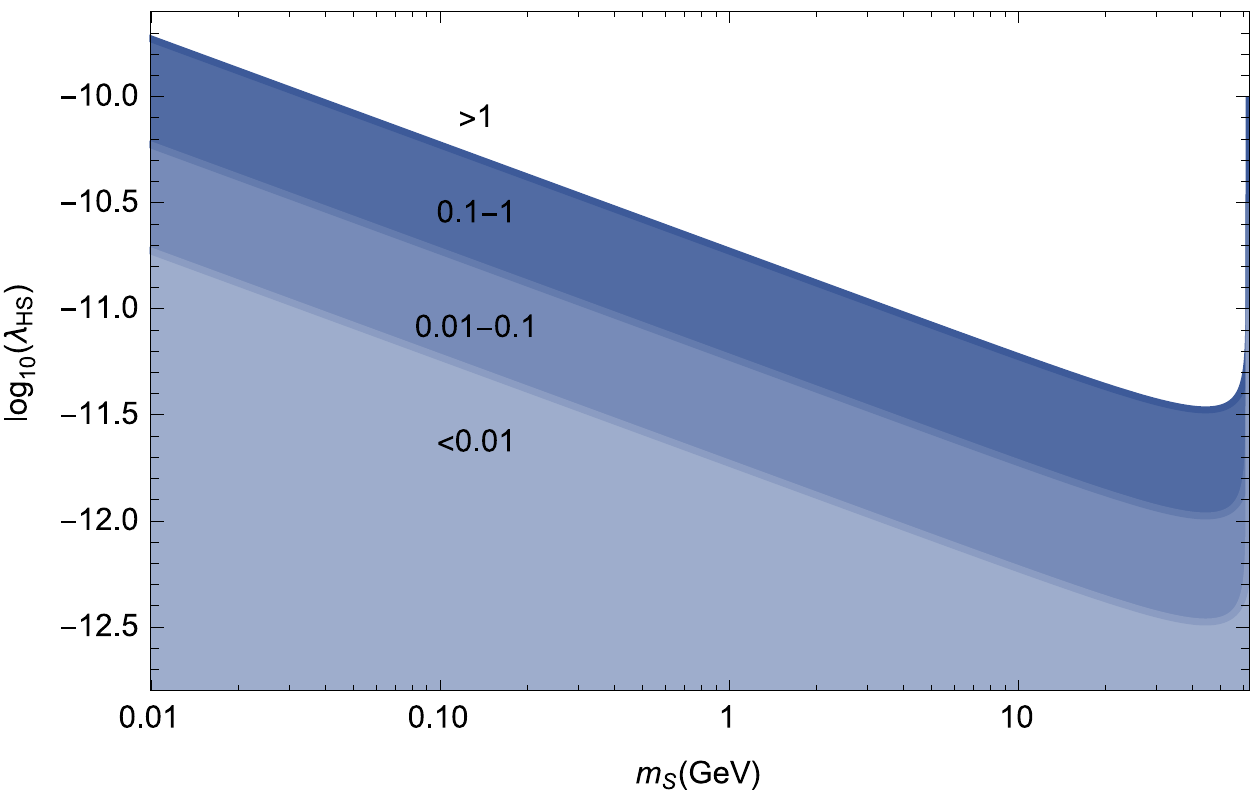}
\caption{The contours, from top to bottom, show the values of $\lHS$ and $m_S$ required to produce relic density $f_{\rm{rel}}=1, 0.1, 0.01$, respectively, via the freeze-in mechanism. The slope of the curves at constant $\Omega_S h^2$ is $1/2$, corresponding to the scaling $\lHS\sim m_S^{-1/2}$; see Eq. (\ref{freezeinlvsm}).}
\label{FIscalar}
\end{center}
\end{figure}

The result on the relic density in the case of freeze-in is shown in Fig. \ref{FIscalar} as a function of the scalar mass, $m_S$, and the portal coupling, $\lHS$.  The contours show the values of $f_{\rm{rel}}$, defined in Eq. (\ref{freldef}), as indicated explicitly in the figure. 

We do not consider the phenomenology of this freeze-in scenario in more detail here, but rather turn to a model where a strong electroweak phase transition can be obtained simultaneously with the observed dark matter relic density. Moreover, the relic density can be produced either via thermal freeze-out or out-of-equilibrium freeze-in scenarios.

\subsection{Fermion dark matter}
\label{fermionDM}

In this section we consider a model where the SM matter content is extended with the singlet sector containing a real scalar and a fermion \cite{Fairbairn:2013uta,Li:2014wia,Fedderke:2014wda}.
The scalar sector is given by the full potential in Eq. (\ref{scalarpotII}), as we are not assuming 
$Z_2$ symmetry for the singlet scalar.
Rather, in this model the dark matter candidate is the singlet fermion, which enters through the Lagrangian
\be
{\cal L}_{\rm{DM}}=\bar{\psi}(\mathrm{i}\slashed\partial-m)\psi+g_S S\bar{\psi}\psi.
\label{fermionlag}
\ee
We assume that the singlet fermion carries an exact global fermion number symmetry or that its mixing with SM neutrinos is otherwise forbidden. The main motivation for this model over the simple scalar case considered in previous section is that since the properties of the electroweak phase transition are affected by the scalar sector, and the dark matter relic density is determined by the properties of the singlet fermion, one can simultaneously explain both. As we will discuss in the following, the dark matter relic density can arise either via freeze-out or freeze-in mechanisms.

At $T=0$, the potential (\ref{scalarpotII}) has several local extrema, which either break or conserve electroweak symmetry. 
There are altogether three symmetric extrema ($v=0$), two of which are minima. We denote one of these by $S=w_0$. Then the two other extrema are at $S=0$ and $S=-w_0-\mu_3/\lambda_S$. 
 Since we do not assume $Z_2$ symmetry for the singlet scalar, the electroweak broken minimum is at $(v = 246$ GeV$,w)$ where the vev of the singlet scalar, $w$, is not necessarily zero. 

We trade the parameters $\mu_H^2$, $\mu_S^2$ and $\muHS$ with $v,w$ and $w_0$ using the extremization conditions
\bea
&\mu_H^2 = \frac{-2 \lambda_H  v^4+\lHS v^2 w^2+4 w^2 (w-w_0) (\mu_3+\lambda_S (w+w_0))}{2 v^2}, \\
&\mu_S^2 = -w_0 (\mu_3+\lambda_S w_0), \\
&\muHS = -\frac{w \left(\lHS v^2+2 (w-w_0) (\mu_3+\lambda_S (w+w_0))\right)}{v^2}.
\eea
In the electroweak symmetric minimum, the requirement that the eigenvalues of the Hessian matrix have to be positive gives
\bea
w_0 (\mu_3+2\lambda_S w_0) &\geq 0, \\
\lHS v^2 + 4 w \left(\mu_3+\lambda_S (w+w_0)\right) &\geq \frac{2 \lambda_H  v^4}{(w-w_0)^2}.
\eea

Moreover, we trade the parameters $w$, $\lambda_H$ and $m$ with physical masses $m_h = 126$ GeV, $m_H$ and $m_{\psi}$ so that finally the free parameters are $m_{\psi},m_H,\lambda_S,\lHS,w_0,\mu_3$ and $g_S$. We perform Monte Carlo scan of the parameter space with
\be
\begin{gathered}
0<\lambda_S<\pi, \quad -2 \sqrt{\lambda_H \lambda_S}<\lHS<\pi, \quad -\pi<g_S<\pi, \\ 
-4000\GeV<\mu_3<4000\GeV, \quad -2000\GeV<w_0<2000\GeV, \\ 0\GeV<m_{\psi}<800\GeV, \quad 200\GeV<m_H<1400\GeV.
\end{gathered}
\ee
In addition to the constraints listed above, we check that the electroweak broken minimum is the deepest one at $T = 0$ and require perturbativity up to scales $\mu\sim 1.5$ TeV. The RG equations for the dimensionless couplings of the model are as follows:
The $\beta$ function of the Yukawa coupling, $g_S$, is
	\begin{align}
	    16\pi^2\beta_{g_S}=&5g_{S}^3,
	\end{align}
	while the $\beta$ functions of the quartic couplings are
	\begin{align}
	    \begin{split}
		16\pi^2 \beta_{\lambda_H}=& 24\lambda_H^2+\frac{1}{2}\lHS^2-3\left(3g^2
		    +g^{\prime 2}-4y_t^2\right)\lambda_H\\
		&+\frac{3}{8}\left(3g^4+2g^2g^{\prime 2}+g^{\prime 4}\right)-6y_t^4, 
	    \end{split} \\
	    \begin{split}
		16\pi^2\beta_{\lHS}=& 4\lHS^2+\left(12\lambda_H+6\lambda_S\right)\lHS\\
		&-3\left(\frac{3}{2}g^2+\frac{1}{2}g^{\prime 2}-2y_t^2-\frac{4}{3}g_S^2\right)\lHS,
	    \end{split} \\
	    \begin{split}
		16\pi^2\beta_{\lambda_{S}}=&18\lambda_S^2+2\lHS^2+8\lambda_S g_S^2-8 g_S^4.
	    \end{split}
	\end{align}
Note that these reproduce the results for the $Z_2$ symmetric scalar case considered in the previous section in the limit 
$g_S\rightarrow 0$.

Since we are not assuming $Z_2$ symmetry for the singlet $S$, fields $S$ and $\phi_\mathrm{r}^0$ are not mass eigenstates. Rather the mass eigenstates are of the form (\ref{scalarmixing}), and the mixing pattern is constrained by the LHC data. As already discussed in Sec. \ref{model}, since the lighter mass eigenstate is identified with the $m_h=126$ GeV boson observed at the LHC, the presence of the singlet component constrains the allowed values of the mixing angle. Due to the simple mixing pattern, all couplings of the lighter mass eigenstate are suppressed by $\cos\beta$ relative to the couplings of the SM Higgs boson. 
We perform a global fit to the current data taking 
signal strengths from the ATLAS, CMS and Tevatron experiments 
\cite{ATLAS-CONF-2014-009, Chatrchyan:2013zna, Chatrchyan:2014nva, CMS:ril, Chatrchyan:2013iaa, Chatrchyan:2013mxa}. We then allow only parameters which are within $2\sigma$ limit, $\cos\beta > 0.85$, of the best fit value, $\cos\beta = 0.95$. We also impose constraints from the precision electroweak measurements on the oblique corrections, i.e. $S$ and $T$ parameters, using formulae given in \cite{Grimus:2008nb}. For the experimental input we use $S=0.04\pm 0.09$ and $T=0.07\pm 0.08$ with correlation of $0.88$ from \cite{Beringer:1900zz}, and accept only points which are within the $2\sigma$ ellipsis around the central value quoted above.
The constraint from the invisible width is similar as in the case of $Z_2$ symmetric scalar. The essential difference is that now there are two channels contributing to the invisible width, $h^0\rightarrow H^0H^0$ and $h^0\rightarrow \psi\psi$,
\be
\Gamma_{h^0\rightarrow H^0H^0}=\frac{\lambda_{hHH}^2}{32\pi m_h}\sqrt{1-\frac{4m_H^2}{m_h^2}},\quad
\Gamma_{h^0\rightarrow\psi\psi}=\frac{g_h^2 m_h}{8\pi} \left(1-\frac{4m_\psi^2}{m_h^2}\right)^\frac{3}{2},
\ee
where $g_h = g_S \sin\beta$ and the coupling $\lambda_{hHH}$ is given by Eq. \eqref{hHHcoup}.
 
To calculate the relic abundance of the singlet fermion, we again apply first the usual freeze-out formalism. The annihilation channels are similar to those of the $Z_2$-symmetric scalar case, as the annihilation to SM fields can again proceed only via the scalar portal interactions. The cross sections are given by the fairly complicated and unilluminating expressions collected in Appendix \ref{xsect}.

The thermal corrections are given by Eqs. (\ref{thermalcorr}) and (\ref{thermalc}), with the additional contributions $g_S^2/6$ and $-g_S m/6$ to $c_S$ and $c_1$, respectively, arising from the coupling between the singlet fermion and scalar.\footnote{Note that the correction to $c_1$ is proportional only to the $m$ term in Eq. (\ref{fermionlag}), and not to the part of the mass term arising from the interaction with $S$. In particular, if the fermion mass would be entirely due to the condensation of $S$, this contribution to $c_1$ would not arise at all.} Using the temperature dependent effective potential we again monitor the evolution of the electroweak symmetric and broken minima as $T$ is increased from $T=0$. The critical temperature, $\Tc$, is obtained from the condition that the symmetric minimum becomes deeper than the broken one. To  check the strength of the phase transition, we evaluate $v(\Tc)/\Tc$.

\begin{figure}[htb]
\begin{center}
\includegraphics[width=0.65\textwidth]{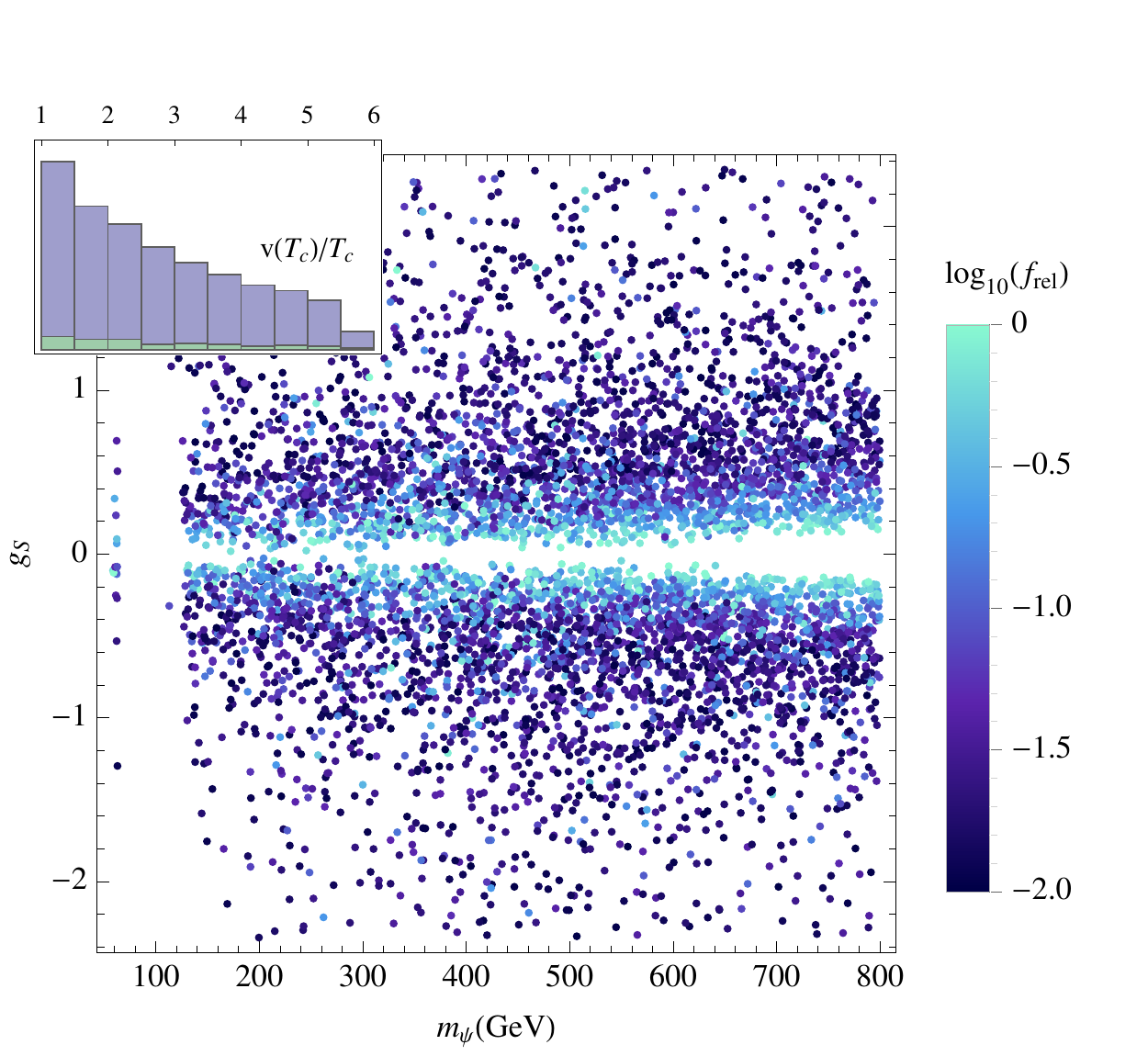}
\caption{Dark matter density as a function of the dark matter mass $m_\psi$ and the Yukawa coupling $g_S$. The inset shows the distribution of values of $v(\Tc)/\Tc$ corresponding to the points in the plot. The shaded lower portion of the histogram bars correspond to points which yield $f_{\rm{rel}}>0.5$.
The constraints from the perturbativity of couplings and LHC invisible Higgs width 
have been taken into account. Also, we show only the points 
which give $0.01 < f_\mathrm{rel} \le 1$, $v(\Tc)/\Tc > 1$ and $\Tc > 40$ GeV. All shown points are also compatible with the LUX constraints.}
\label{constraints3}
\end{center}
\end{figure}

The results on the dark matter abundance are shown in Fig. \ref{constraints3} as a function of the mass of the WIMP candidate, $m_\psi$, and its coupling with the singlet scalar, $g_S$. We see that a sizable fraction of the dark matter relic abundance can be realized with practically any mass larger than $m_h$, but the viable values of $g_S$ are confined within narrow interval, $g_S\simeq {\cal O}(0.1)$.  On the other hand, the model now easily produces strong electroweak phase transition. This is shown in the inset of the left panel of Fig. \ref{constraints3}, where the shaded lower (green) parts of the histogram correspond to points which yield $f_{\rm{rel}}>0.5$. 

All the points shown in Fig. \ref{constraints3} are compatible with the direct search constraint from LUX experiment. To compare with the direct searches, we again evaluate the spin-independent WIMP-nucleon cross section. Now, the interaction between the WIMP and nucleons contains the factors
\be
\tilde{g}_h=g_S\sin\beta, \; \tilde{g}_H =g_S\cos\beta,
\ee
which leads to
\be
\sigma_{\rm{SI}}^0= \frac{\mu_N^2 f_N^2 m_N^2}{\pi v^2} g_S^2 \sin^2\beta\cos^2\beta \left(\frac{1}{m_h^2} - \frac{1}{m_H^2} \right)^2.
\ee

\begin{figure}[htb]
\begin{center}
\includegraphics[width=0.49\textwidth]{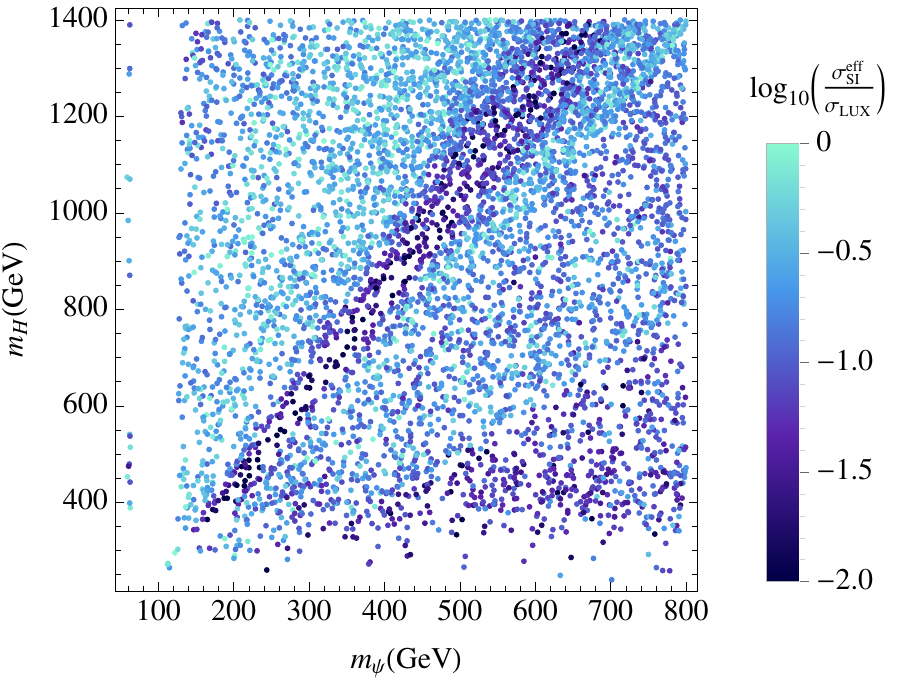} \hspace{.1cm}
\includegraphics[width=0.47\textwidth]{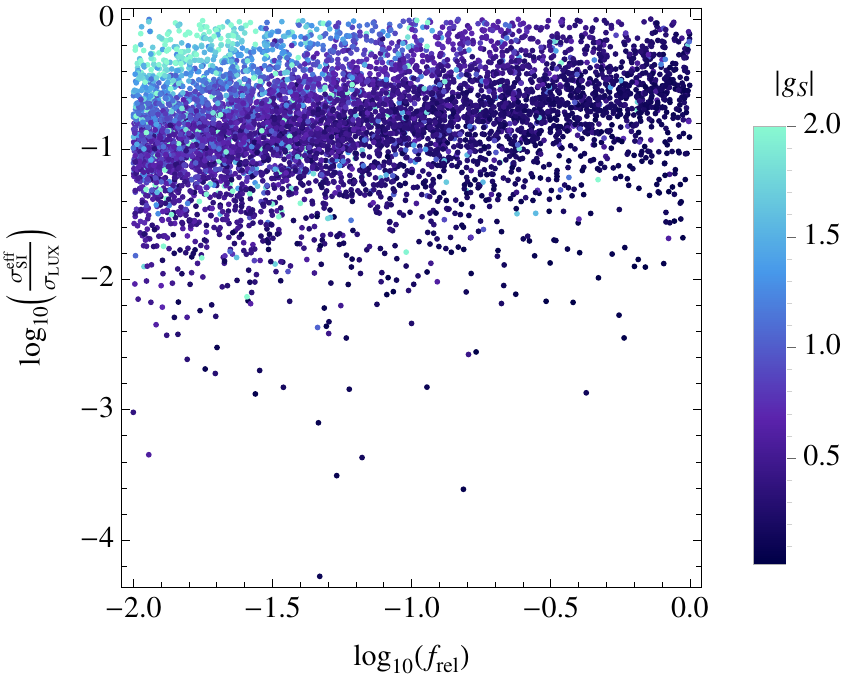}
\caption{Left panel: the same data points as in Fig. \ref{constraints3}  in the $(m_\psi,m_H)$ plane. The color code now shows the magnitude of relative cross section $\sigma_{\rm{SI}}^{\rm{eff}}/\sigma_{\rm{LUX}}$.
Right panel: the data points as a function of $f_{\rm{rel}}$ and 
$\sigma_{\rm{SI}}^{\rm{eff}}/\sigma_{\rm{LUX}}$. The color coding corresponds to the magnitude of the coupling $g_S$. 
In both panels the constraints from the perturbativity of couplings and LHC invisible Higgs width 
have been taken into account. Also, we show only the points 
which give $0.01 < f_\mathrm{rel} \le 1$, $v(\Tc)/\Tc > 1$ and $\Tc > 40$ GeV and which are not excluded by the LUX data.}
\label{constraints4}
\end{center}
\end{figure}

A detailed account of the direct search limits from LUX experiment are shown in Fig.~\ref{constraints4}. The left panel shows the points from Fig. \ref{constraints3} in the $(m_\psi,m_H)$ plane, with the color coding now corresponding to the magnitude of the relative cross section $\sigma_{\rm{SI}}^{\rm{eff}}/\sigma_{\rm{LUX}}$. We see that the points furthest below the LUX bound are mostly concentrated around the region $m_H\sim 2m_\psi$. The right panel of Fig. \ref{constraints4} shows the distribution of the results with respect to the values of $f_{\rm{rel}}$ and $\sigma_{\rm{SI}}^{\rm{eff}}/\sigma_{\rm{LUX}}$, with the color coding now corresponding to the values of $g_S$. Here we see the result already observed in Fig. \ref{constraints3} that the large values of the relic density are confined to values of $g_S\sim{\cal O}(0.1)$ and that, in addition, the compatibility with the LUX bound is controlled by the mixing pattern in the scalar sector.

\begin{figure}
\begin{center}
\includegraphics[width=0.6\textwidth]{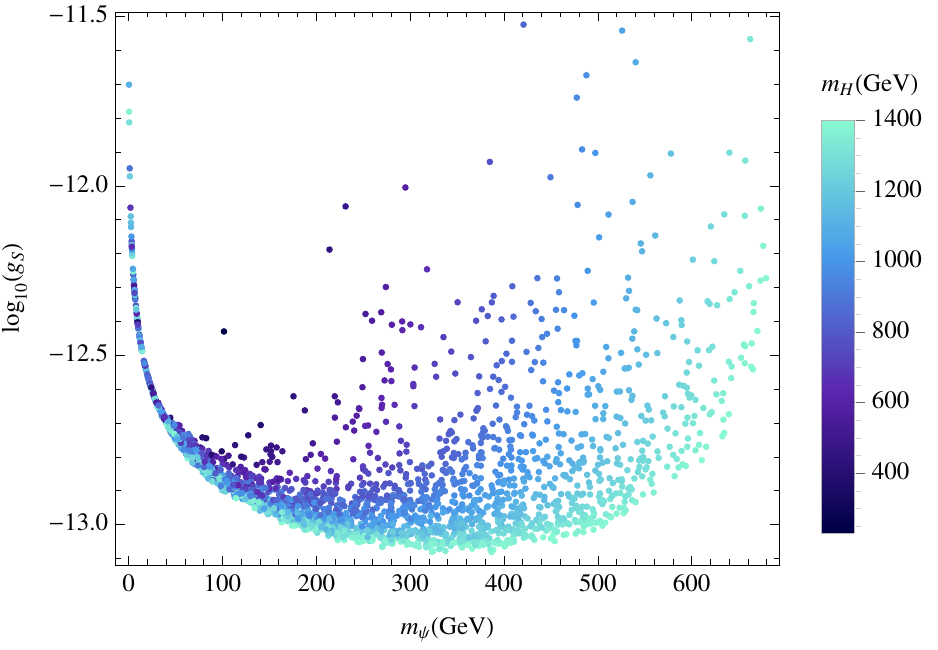}
\caption{The relic density $f_{\rm{rel}}=1$ produced via freeze-in mechanism as a function of dark matter mass $m_\psi$ and Yukawa coupling $g_S$. For all points $f_{\rm{rel}} = 1$, $v(\Tc)/\Tc > 1$ and $\Tc > 40$ GeV.}
\label{FIfermion}
\end{center}
\end{figure}

To conclude this phenomenological analysis, let us again consider the dark matter relic density in the case of freeze in. Now the dark matter number density is produced from $h^0$ and $H^0$, which both contain the component of the singlet $S$ which couples to the dark matter candidate $\psi$. In Fig. \ref{FIfermion} we show the values of mass $m_\psi$ and coupling $g_S$ for which $\Omega_\psi h^2=0.12$ can be obtained. Of course, with such weak coupling the model in this part of the parameter space remains unconstrained by the dark matter direct search experiments. 
At all points shown in Fig. \ref{FIfermion}, also a strong electroweak phase transition can be realized. This is possible since the relic density is determined by the coupling $g_S$ while the properties of the phase transition are controlled by the couplings in the scalar sector. Similarly to the scalar case considered earlier, also here the characteristic feature of the mechanism is superweak coupling, $\Omega_\psi h^2\sim 10^{-12}\sqrt{\mathrm{GeV}/m_\psi}$, and the region where the mechanism can be applied is bounded by $m_\psi=m_H/2$.

\subsection{Stability}

One of the intriguing properties of the Standard Model with a light Higgs, $m_h=126$ GeV, is that it is very near of being stable up to the Planck scale. In the SM the Higgs self-coupling becomes negative at one-loop level around energies $\mu\simeq 10^{8}$ GeV, but only so slightly that the theory remains in the region of metastability.\footnote{A higher-order analysis shows that this metastability scale is actually a little bit higher in the SM, around $\mu\simeq 10^{11}$ GeV~\cite{Degrassi:2012ry,Antipin:2013sga}, but a one-loop analysis is sufficient for our purposes here.} 
It is well known that extra singlets affect this situation, see e.g. \cite{Haba:2014sia}. For example, in our model we find that setting $\lambda_S(m_t)=g_S(m_t)=0$, the Higgs self-coupling remains positive for $\lHS(m_t)\gtrsim 0.4$ all the way up to the Planck scale. Currently there is some motivation for such stabilization from the measurement of the polarization of the cosmic microwave background by the BICEP-2 experiment \cite{Ade:2014xna}. The observation seems to be consistent with the tensor-scalar ratio $r=0.16^{+0.06}_{-0.05}$ which, if due to gravitational waves produced during inflation, sets the scale of the energy density to be very large, of the order of $(10^{16}\GeV)^4$, during inflation. This makes the computations of metastability with respect to quantum tunneling futile, since such large energy densities allow the Higgs field to classically roll into the instability; see e.g. \cite{Espinosa:2007qp,Kobakhidze:2013tn,Spencer-Smith:2014woa}. Possible solutions arising from modifying the inflationary dynamics were recently investigated in \cite{Fairbairn:2014zia}. Of course another obvious solution arises if the beyond SM degrees of freedom force the scalar couplings to guarantee positivity of the potential at large field values up to the Planck scale. Let us, therefore, investigate the possibilities further in the model we have studied here.

\begin{figure}
\begin{center}
\includegraphics[width=0.46\textwidth]{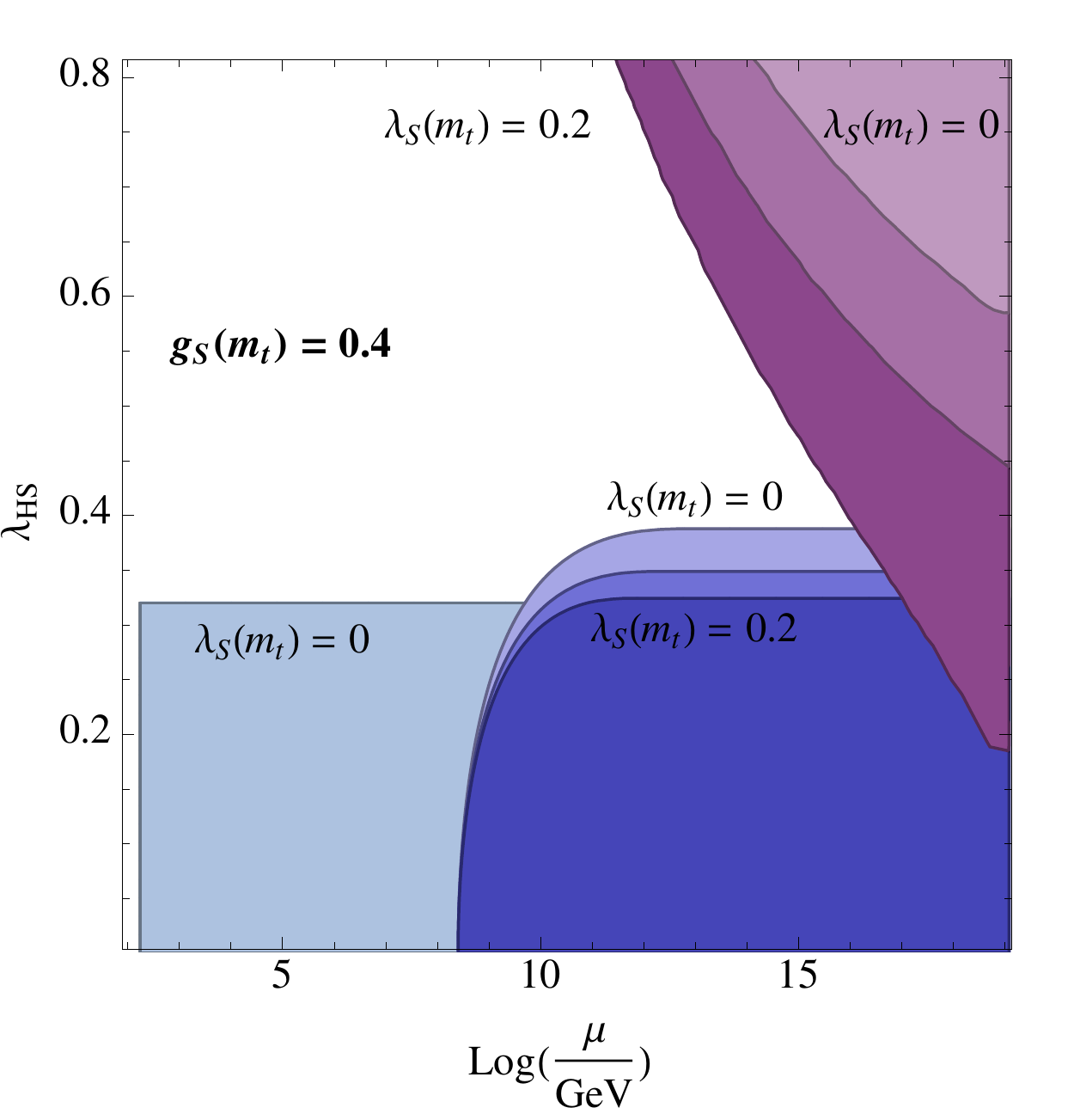} \hspace{0.4cm}
\includegraphics[width=0.46\textwidth]{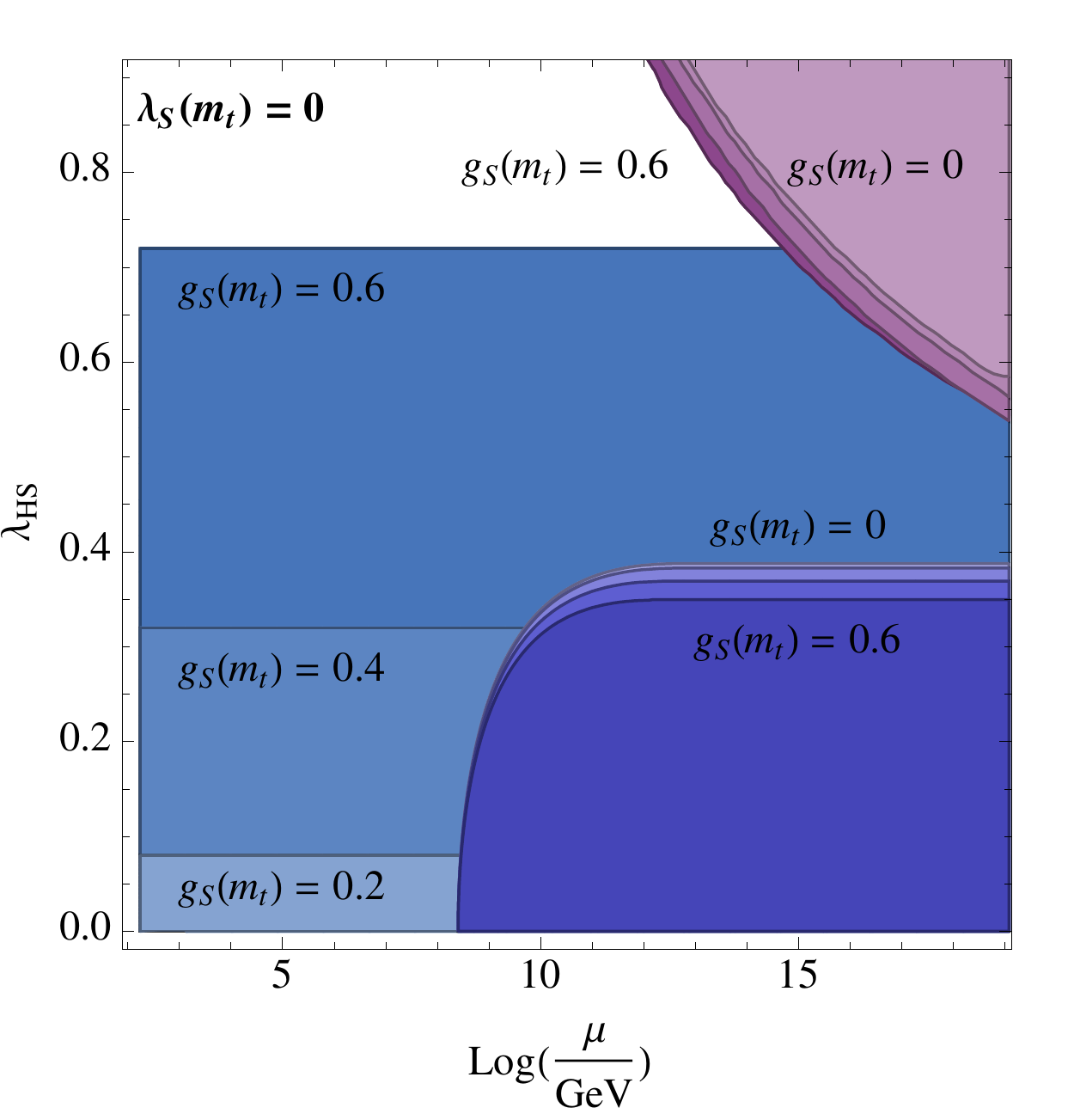}
\caption{The figures show constraints from vacuum stability and perturbativity of the couplings. The contours in the lower right corner show the regions where the Higgs self-coupling becomes negative, while the contours in the upper right corner show the regions where one or more of the couplings become large. Finally, the horizontal contours correspond to $\lambda_S$ becoming negative. In the left panel $g_s(m_t)=0.4$, and the contours show the dependence on $\lambda_S(m_t)$, while the right panel shows the dependence on $g_s(m_t)$ at $\lambda_S(m_t)=0$.}
\label{stability}
\end{center}
\end{figure}

The results are shown in Fig. \ref{stability}. The shaded regions in the lower right corner correspond to parameter domains where the Higgs self-coupling becomes negative and signal the vacuum instability. The shaded regions in the upper right corner show the parameter domains where some of the couplings become non-perturbative. Finally, the horizontal lines show the values where the coupling $\lambda_S$ is driven negative. For simplicity, the results of Fig.~\ref{stability} depict the $Z_2$-symmetric case for which the Higgs self-coupling has the SM value at the electroweak scale, $\lambda_H(m_t)=0.128$. The left panel of Fig. \ref{stability} shows the dependence on $\lambda_S(m_t)$ at value of $g_S(m_t)=0.4$. The figure shows, in particular, how sensitive the perturbativity of the model is on the value of $\lambda_S$: if perturbativity is to be required all the way to the Planck scale, we must have $\lambda_S\lesssim 0.2$. The right panel of Fig. \ref{stability} illustrates similarly the dependence on $g_S(m_t)$: increasing $g_S(m_t)$ above ca. 0.5 forces an instability on $\lambda_S$ unless $\lHS$ becomes large, which in turn forces the theory into a non-perturbative domain.

These results lead to a narrow range of couplings which correspond to a model which remains perturbative and stable up to the Planck scale. These ranges are $\lambda_S(m_t)\lesssim 0.2$, $g_S(m_t)\lesssim 0.6$ and $0.35\lesssim \lHS\lesssim 0.55$.

A further effect on the stability is obtained by relaxing the $Z_2$ symmetry of the potential. Although allowing 
trilinear terms in the scalar potential does not affect the $\beta$~functions of the quartic couplings, it does change the 
vacuum structure thereby mixing the singlet and the doublet scalars at electroweak scale. This in turn changes the value of the 
coupling $\lambda_H$ at the electroweak scale when compared with the SM value. We do a random scan over the parameter space
to illustrate that within our model it is possible to
obtain a considerable amount of the dark matter and simultaneously to have a strong electroweak phase transition with a stable 
potential all the way up to the Planck scale. As a result, we find a benchmark point with 
\be
\lambda_H(m_t)=0.197,\,\,  \lambda_S(m_t)=0.053,\,\,
\lHS(m_t)=0.376,\,\, g_S(m_t)=0.247,
\label{benchmarkpoint}
\ee
 for which $f_{\mathrm{rel}}=0.94$ and $v(\Tc)/\Tc=1.3$. 

Finally, let us discuss possible implications for the LHC. The above benchmark point serves as a concrete example. At this point the singlet fermion mass is $m_\psi=679$ GeV and the heavy scalar mass is $m_H=371$ GeV. Both of these are above $m_h/2$ so the hidden sector will remain invisible with zero branching fraction. However, even if the new degrees of freedom are not expected to manifest at LHC, the model predicts deviations with respect to SM. As already mentioned above, the mixing of the singlet and doublet scalars affects the coupling $\lambda_H$, and this is directly connected with the scalar potential. In the SM the Higgs self-coupling has value $\lambda_{\rm{SM}}(m_t)=0.128$, while for example for the parameter space point in Eq. (\ref{benchmarkpoint}) we have $\lambda_H(m_t)=0.197=1.5\lambda_{\rm{SM}}(m_t)$. 

\begin{figure}
\begin{center}
\includegraphics[width=0.5\textwidth]{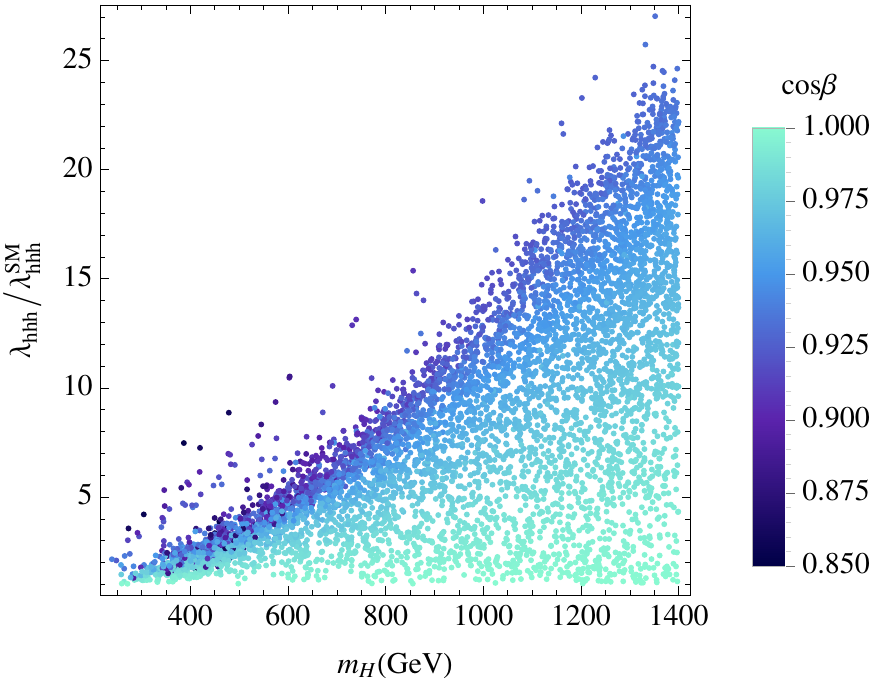} \hspace{0.2cm}
\includegraphics[width=0.4\textwidth]{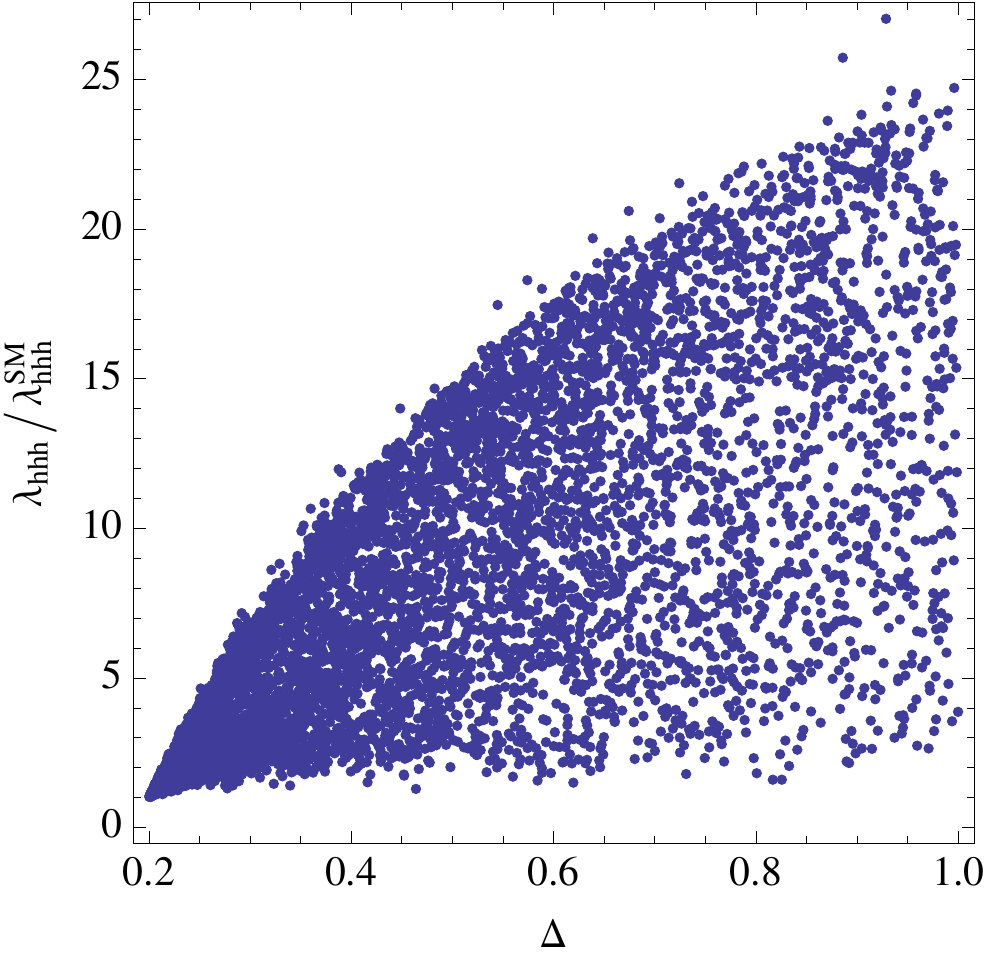}
\caption{Left panel: the value of the trilinear Higgs coupling, $\lambda_{hhh}$, in the model as a function of $m_H$ and $\cos\beta$. The points correspond to the ones shown in Fig. \ref{constraints3}. Right panel: the value of the trilinear Higgs coupling as a function of the deviation $\Delta$ from the optimal values of the fit to the $S$ and $T$ parameters. The value $\Delta=1$ corresponds to the $2\sigma$ limit.}
\label{threehiggscoupling}
\end{center}
\end{figure}

To search for deviations from the SM all the Higgs couplings should be measured and compared with the predicted SM values. The cubic coupling of three physical Higgs bosons, $\lambda_{hhh}$, could be measured at LHC in the production of two Higgs bosons, see e.g. \cite{Barger:2013jfa}. For the benchmark point, 
Eq. (\ref{benchmarkpoint}), we have $\lambda_{hhh}/\lambda_{hhh}^{\rm{SM}}=2.00$.
Generally, the dependence of $\lambda_{hhh}/\lambda_{hhh}^{\rm{SM}}$ on $m_S$ and singlet-doublet mixing $\cos\beta$ within the allowed $2\sigma$ range in the model is shown in the left panel of Fig. \ref{threehiggscoupling}. We find that sizable deviations from the SM value are indeed expected. The growth of the deviation can be traced to the dimensionful couplings, in particular to $\mu_3$, whose values can be large in comparison to the electroweak scale, $v$. On the other hand, the large intrinsic scales will also affect the $S$ and $T$ parameters through the mixing of singlet and doublet scalars. This is illustrated by the right panel of Fig. \ref{threehiggscoupling}, which shows the trilinear Higgs coupling with respect to the deviation, $\Delta$, from the best fit value of the oblique parameters $S$ and $T$, the value $\Delta=1$ corresponding to the $2\sigma$ limit. All our data points are naturally within this constraint by construction.

\section{Conclusions}
\label{checkout}

In this paper we have considered simple singlet extensions of the SM motivated by the possibility to explain cosmological paradigms of dark matter and electroweak baryogenesis. For the latter we have focused on the possibility to realize a strong electroweak phase transition which is a necessary general condition for any particular realization of creating the matter-antimatter asymmetry.

We considered two concrete realizations: First we studied a model where the dark matter is a singlet scalar protected by a discrete $Z_2$ symmetry in the zero temperature vacuum. Second, we considered a model where the dark matter candidate is a singlet fermion coupled with the SM via a singlet scalar. In this case the scalar potential can be arranged to lead to strong electroweak phase transition while the singlet fermion saturates the observed dark matter abundance. In this case, the dark matter abundance can be produced either via freeze-out or freeze-in mechanisms.

Between these two models, the main difference which is relevant for the phenomenological aspects of dark matter and strong first-order phase transition is that in the $Z_2$-symmetric-scalar case both aspects arise from the same field, whereas in the fermion case the dark matter abundance is due to the fermion while the modifications to the electroweak transition are entirely due to the portal scalar. One can also imagine alternative realizations based on similar dichotomy. For example, one could consider two different scalar fields. Imposing a discrete symmetry on only one of them would make the field a plausible origin of dark matter similar to the $Z_2$-symmetric case we considered here, while the other field would provide the necessary modifications to strengthen the electroweak phase transition. 

We also studied the implications of the RG equations on the stability and perturbativity of the model. We found that due to the singlet scalar, the Higgs self-coupling can remain positive at all energies below the Planck scale. With suitable values of the portal coupling, $\lHS$, the self-coupling of the singlet scalar, $\lambda_S$, and the scalar Yukawa coupling, $g_S$, the entire model was found to remain perturbative and stable all the way up to the Planck scale. 

A possible experimental access to the singlet sector is provided via the couplings of the Higgs boson, in particular the trilinear coupling of three physical Higgs bosons.  
If measured at LHC, and deviations from the corresponding value in the SM are found, this can provide, if no new states are directly observed at the LHC experiments, important clues about a possible hidden sector akin to the models studied in this paper. 

\acknowledgments This work has been financially supported by Finnish Cultural Foundation and by the Academy of Finland project 267842.

\appendix

\section{Cross sections}
\label{xsect}

Here we give the formulae for the computation of the annihilation cross section for the model considered in Sec. \ref{fermionDM}. To make the equations more concise, it is useful to define the couplings
\begin{align}
\begin{split}
\lambda_{hhh} &= -6 \lambda_H  v \cos^3\beta-3 \lHS v \sin^2\beta \cos\beta - 2 \sin^3\beta (\mu_3+3 \lambda_S w) \\
&- 3 \sin\beta \cos^2\beta (\mu_H+\lHS w), 
\end{split} \\
\begin{split}
\lambda_{hhH} &= -2 v \sin\beta \cos^2\beta (\lHS-3 \lambda_H ) + \lHS v \sin^3\beta \\
&+\sin\beta \sin (2 \beta ) (-\mu_3+\mu_H-3 \lambda_S w+\lHS w) -\cos^3\beta (\mu_H+\lHS w), 
\end{split}\\
\begin{split} \label{hHHcoup}
\lambda_{hHH} &= v \sin\beta \sin (2 \beta ) (\lHS-3 \lambda_H )-\lHS v \cos^3\beta \\
&- 2 \sin\beta \cos^2\beta (\mu_3-\mu_H+3 \lambda_S w-\lHS w) - \sin^3\beta (\mu_H+\lHS w), 
\end{split}\\
\begin{split}
\lambda_{HHH} &= 6 v \sin^2\beta \cos\beta (\lHS-3 \lambda_H )-3 \lHS v \cos^3\beta \\
&- 6 \sin\beta \cos^2\beta (\mu_3-\mu_H+3 \lambda_S w-\lHS w) - 3 \sin^3\beta (\mu_H+\lHS w), 
\end{split}
\end{align}
and
\bea
&g_h = g_S \sin\beta, \quad g_H = g_S \cos\beta, \quad Y_h = \frac{m_f}{v}\cos\beta, \quad Y_H = -\frac{m_f}{v}\sin\beta, \\
&g_{hZ} = \frac{v (g^2 + g'^2)}{2} \cos\beta, \quad g_{hW} = \frac{v g^2}{2} \cos\beta, \\ 
&g_{HZ} = -\frac{v (g^2 + g'^2)}{2} \sin\beta, \quad g_{HW} = -\frac{v g^2}{2} \sin\beta.
\eea

The squared amplitude, averaged over the initial states and summed over the final states, for the dark matter, $\psi$, annihilating to two scalars, $h_i,h_j = h^0,H^0$, is given by
\bea
\left|T(\psi\overline\psi\to h_i h_j)\right|^2 &= \left(1-\frac{\delta_{ij}}{2}\right) \left( 2(s-4m_{\psi}^2) \left| \sum_{k=h,H} \frac{g_k \lambda_{ijk}}{s-m_k^2+\i m_k \Gamma_k} \right|^2 \right. \\
&+2g_i^2 g_j^2 \left( tu -m_{\psi}^2(t+u) + m_{\psi}^4 - m_i^2 m_j^2 \right)  \left( \frac{1}{t-m_{\psi}^2} - \frac{1}{u-m_{\psi}^2} \right)^2 \\
& \left. +4 m_{\psi} (t-u) \left( \frac{1}{t-m_{\psi}^2} - \frac{1}{u-m_{\psi}^2} \right)\sum_{k = h,H} \frac{g_i g_j g_k \lambda_{ijk}(s-m_k^2)}{\left|s-m_k^2+\i m_k \Gamma_k\right|^2} \right) .
\eea
Similarly the squared amplitudes, averaged over the initial states and summed over the final states, to fermion and gauge boson final states are
\be
\left|T(\psi\overline\psi\to ff)\right|^2 = s^2 v_{\psi}^2 v_f^2X_f\left| \sum_{k=h,H} \frac{g_k Y_k}{s-m_k^2+\i m_k \Gamma_k} \right|^2,
\ee
and
\be
\left|T(\psi\overline\psi\to VV)\right|^2 = 2 s v_{\psi}^2 \left( 3-\frac{s}{M_V^2} + \frac{s^2}{4M_V^4} \right) \left| \sum_{k=h,H} \frac{g_k g_{kV}}{s-m_k^2+\i m_k \Gamma_k} \right|^2 \delta_V.
\ee
In the expression for the fermion channel the factor $X_f=1$ for leptons in the final state, while for quarks
\be
X_f=3\left(1+\left(\frac{3}{2}\ln\frac{m_q^2}{s}+\frac{9}{4}\right)\frac{4\alpha_s}{3\pi}\right),
\ee
where $\alpha_s=0.12$ is the strong coupling constant. Similarly as in the scalar case \cite{Cline:2013gha}, also for fermion dark matter the region where this correction is important is ruled out by the invisible width of the Higgs.

Then, the cross section for the process $\psi\overline\psi\to ij$ reads
\be
\sigma(\psi\psi\to ij) = \frac{1}{16 \pi s^2 v_{\psi}^2} \int_{t_-^{(ij)}}^{t_+^{(ij)}} \mathrm{d}t \left|T(\psi\psi\to ij)\right|^2,
\ee
where
\be
t_\pm^{(ij)} = m_{\psi}^2 + \frac{1}{2}(m_i^2+m_j^2-s) + \frac{1}{2}\sqrt{\left(1-\frac{4m_{\psi}^2}{s}\right)(s^2+(m_i^2-m_j^2)^2-2s(m_i^2+m_j^2))}.
\ee

\bibliography{sm+singlets_JCAP}

\providecommand{\href}[2]{#2}\begingroup\raggedright\begin{thebibliography}{10}

\bibitem{Carmi:2012yp}
D.~Carmi, A.~Falkowski, E.~Kuflik, and T.~Volansky, {\it {Interpreting LHC
  Higgs Results from Natural New Physics Perspective}},  {\em JHEP} {\bf 1207}
  (2012) 136, [\href{http://xxx.lanl.gov/abs/1202.3144}{{\tt
  arXiv:1202.3144}}].

\bibitem{Espinosa:2012ir}
J.~Espinosa, C.~Grojean, M.~Muhlleitner, and M.~Trott, {\it {Fingerprinting
  Higgs Suspects at the LHC}},  {\em JHEP} {\bf 1205} (2012) 097,
  [\href{http://xxx.lanl.gov/abs/1202.3697}{{\tt arXiv:1202.3697}}].

\bibitem{Giardino:2012ww}
P.~P. Giardino, K.~Kannike, M.~Raidal, and A.~Strumia, {\it {Reconstructing
  Higgs boson properties from the LHC and Tevatron data}},  {\em JHEP} {\bf
  1206} (2012) 117, [\href{http://xxx.lanl.gov/abs/1203.4254}{{\tt
  arXiv:1203.4254}}].

\bibitem{Alanne:2013dra}
T.~Alanne, S.~Di~Chiara, and K.~Tuominen, {\it {LHC Data and Aspects of New
  Physics}},  {\em JHEP} {\bf 1401} (2014) 041,
  [\href{http://xxx.lanl.gov/abs/1303.3615}{{\tt arXiv:1303.3615}}].

\bibitem{Aad:2012tfa}
{\bf ATLAS} Collaboration, G.~Aad et~al., {\it {Observation of a new particle
  in the search for the Standard Model Higgs boson with the ATLAS detector at
  the LHC}},  {\em Phys.Lett.} {\bf B716} (2012) 1--29,
  [\href{http://xxx.lanl.gov/abs/1207.7214}{{\tt arXiv:1207.7214}}].

\bibitem{Chatrchyan:2012ufa}
{\bf CMS} Collaboration, S.~Chatrchyan et~al., {\it {Observation of a new boson
  at a mass of 125 GeV with the CMS experiment at the LHC}},  {\em Phys.Lett.}
  {\bf B716} (2012) 30--61, [\href{http://xxx.lanl.gov/abs/1207.7235}{{\tt
  arXiv:1207.7235}}].

\bibitem{McDonald:1993ex}
J.~McDonald, {\it {Gauge singlet scalars as cold dark matter}},  {\em
  Phys.Rev.} {\bf D50} (1994) 3637--3649,
  [\href{http://xxx.lanl.gov/abs/hep-ph/0702143}{{\tt hep-ph/0702143}}].

\bibitem{Burgess:2000yq}
C.~Burgess, M.~Pospelov, and T.~ter Veldhuis, {\it {The Minimal model of
  nonbaryonic dark matter: A Singlet scalar}},  {\em Nucl.Phys.} {\bf B619}
  (2001) 709--728, [\href{http://xxx.lanl.gov/abs/hep-ph/0011335}{{\tt
  hep-ph/0011335}}].

\bibitem{LopezHonorez:2006gr}
L.~Lopez~Honorez, E.~Nezri, J.~F. Oliver, and M.~H. Tytgat, {\it {The Inert
  Doublet Model: An Archetype for Dark Matter}},  {\em JCAP} {\bf 0702} (2007)
  028, [\href{http://xxx.lanl.gov/abs/hep-ph/0612275}{{\tt hep-ph/0612275}}].

\bibitem{Alvares:2012qv}
J.~Ruiz-Alvarez, C.~de~S.~Pires, F.~S. Queiroz, D.~Restrepo, and
  P.~Rodrigues~da Silva, {\it {On the Connection of Gamma-Rays, Dark Matter and
  Higgs Searches at LHC}},  {\em Phys.Rev.} {\bf D86} (2012) 075011,
  [\href{http://xxx.lanl.gov/abs/1206.5779}{{\tt arXiv:1206.5779}}].

\bibitem{LopezHonorez:2012kv}
L.~Lopez-Honorez, T.~Schwetz, and J.~Zupan, {\it {Higgs portal, fermionic dark
  matter, and a Standard Model like Higgs at 125 GeV}},  {\em Phys.Lett.} {\bf
  B716} (2012) 179--185, [\href{http://xxx.lanl.gov/abs/1203.2064}{{\tt
  arXiv:1203.2064}}].

\bibitem{Fairbairn:2013uta}
M.~Fairbairn and R.~Hogan, {\it {Singlet Fermionic Dark Matter and the
  Electroweak Phase Transition}},  {\em JHEP} {\bf 1309} (2013) 022,
  [\href{http://xxx.lanl.gov/abs/1305.3452}{{\tt arXiv:1305.3452}}].

\bibitem{Alves:2013tqa}
A.~Alves, S.~Profumo, and F.~S. Queiroz, {\it {The dark $Z^{'}$ portal: direct,
  indirect and collider searches}},  {\em JHEP} {\bf 1404} (2014) 063,
  [\href{http://xxx.lanl.gov/abs/1312.5281}{{\tt arXiv:1312.5281}}].

\bibitem{Hambye:2008bq}
T.~Hambye, {\it {Hidden vector dark matter}},  {\em JHEP} {\bf 0901} (2009)
  028, [\href{http://xxx.lanl.gov/abs/0811.0172}{{\tt arXiv:0811.0172}}].

\bibitem{Davoudiasl:2013jma}
H.~Davoudiasl and I.~M. Lewis, {\it {Dark Matter from Hidden Forces}},  {\em
  Phys.Rev.} {\bf D89} (2014) 055026,
  [\href{http://xxx.lanl.gov/abs/1309.6640}{{\tt arXiv:1309.6640}}].

\bibitem{Kuzmin:1985mm}
V.~Kuzmin, V.~Rubakov, and M.~Shaposhnikov, {\it {On the Anomalous Electroweak
  Baryon Number Nonconservation in the Early Universe}},  {\em Phys.Lett.} {\bf
  B155} (1985) 36.

\bibitem{Kajantie:1996mn}
K.~Kajantie, M.~Laine, K.~Rummukainen, and M.~E. Shaposhnikov, {\it {Is there a
  hot electroweak phase transition at m(H) larger or equal to m(W)?}},  {\em
  Phys.Rev.Lett.} {\bf 77} (1996) 2887--2890,
  [\href{http://xxx.lanl.gov/abs/hep-ph/9605288}{{\tt hep-ph/9605288}}].

\bibitem{Rummukainen:1998as}
K.~Rummukainen, M.~Tsypin, K.~Kajantie, M.~Laine, and M.~E. Shaposhnikov, {\it
  {The Universality class of the electroweak theory}},  {\em Nucl.Phys.} {\bf
  B532} (1998) 283--314, [\href{http://xxx.lanl.gov/abs/hep-lat/9805013}{{\tt
  hep-lat/9805013}}].

\bibitem{Profumo:2007wc}
S.~Profumo, M.~J. Ramsey-Musolf, and G.~Shaughnessy, {\it {Singlet Higgs
  phenomenology and the electroweak phase transition}},  {\em JHEP} {\bf 0708}
  (2007) 010, [\href{http://xxx.lanl.gov/abs/0705.2425}{{\tt
  arXiv:0705.2425}}].

\bibitem{Espinosa:2011ax}
J.~R. Espinosa, T.~Konstandin, and F.~Riva, {\it {Strong Electroweak Phase
  Transitions in the Standard Model with a Singlet}},  {\em Nucl.Phys.} {\bf
  B854} (2012) 592--630, [\href{http://xxx.lanl.gov/abs/1107.5441}{{\tt
  arXiv:1107.5441}}].

\bibitem{McDonald:2001vt}
J.~McDonald, {\it {Thermally generated gauge singlet scalars as selfinteracting
  dark matter}},  {\em Phys.Rev.Lett.} {\bf 88} (2002) 091304,
  [\href{http://xxx.lanl.gov/abs/hep-ph/0106249}{{\tt hep-ph/0106249}}].

\bibitem{Cline:2012hg}
J.~M. Cline and K.~Kainulainen, {\it {Electroweak baryogenesis and dark matter
  from a singlet Higgs}},  {\em JCAP} {\bf 1301} (2013) 012,
  [\href{http://xxx.lanl.gov/abs/1210.4196}{{\tt arXiv:1210.4196}}].

\bibitem{Cline:2013gha}
J.~M. Cline, K.~Kainulainen, P.~Scott, and C.~Weniger, {\it {Update on scalar
  singlet dark matter}},  {\em Phys.Rev.} {\bf D88} (2013) 055025,
  [\href{http://xxx.lanl.gov/abs/1306.4710}{{\tt arXiv:1306.4710}}].

\bibitem{Li:2014wia}
T.~Li and Y.-F. Zhou, {\it {Strongly first order phase transition in the
  singlet fermionic dark matter model after LUX}},
  \href{http://xxx.lanl.gov/abs/1402.3087}{{\tt arXiv:1402.3087}}.

\bibitem{Aprile:2011hi}
{\bf XENON100} Collaboration, E.~Aprile et~al., {\it {Dark Matter Results from
  100 Live Days of XENON100 Data}},  {\em Phys.Rev.Lett.} {\bf 107} (2011)
  131302, [\href{http://xxx.lanl.gov/abs/1104.2549}{{\tt arXiv:1104.2549}}].

\bibitem{Akerib:2013tjd}
{\bf LUX} Collaboration, D.~Akerib et~al., {\it {First results from the LUX
  dark matter experiment at the Sanford Underground Research Facility}},  {\em
  Phys.Rev.Lett.} {\bf 112} (2014) 091303,
  [\href{http://xxx.lanl.gov/abs/1310.8214}{{\tt arXiv:1310.8214}}].

\bibitem{Duda:2001ae}
G.~Duda, G.~Gelmini, and P.~Gondolo, {\it {Detection of a subdominant density
  component of cold dark matter}},  {\em Phys.Lett.} {\bf B529} (2002)
  187--192, [\href{http://xxx.lanl.gov/abs/hep-ph/0102200}{{\tt
  hep-ph/0102200}}].

\bibitem{Profumo:2009tb}
S.~Profumo, K.~Sigurdson, and L.~Ubaldi, {\it {Can we discover multi-component
  WIMP dark matter?}},  {\em JCAP} {\bf 0912} (2009) 016,
  [\href{http://xxx.lanl.gov/abs/0907.4374}{{\tt arXiv:0907.4374}}].

\bibitem{Aoki:2012ub}
M.~Aoki, M.~Duerr, J.~Kubo, and H.~Takano, {\it {Multi-Component Dark Matter
  Systems and Their Observation Prospects}},  {\em Phys.Rev.} {\bf D86} (2012)
  076015, [\href{http://xxx.lanl.gov/abs/1207.3318}{{\tt arXiv:1207.3318}}].

\bibitem{Lee:1977ua}
B.~W. Lee and S.~Weinberg, {\it {Cosmological Lower Bound on Heavy Neutrino
  Masses}},  {\em Phys.Rev.Lett.} {\bf 39} (1977) 165--168.

\bibitem{Gondolo:1990dk}
P.~Gondolo and G.~Gelmini, {\it {Cosmic abundances of stable particles:
  Improved analysis}},  {\em Nucl.Phys.} {\bf B360} (1991) 145--179.

\bibitem{Ade:2013zuv}
{\bf Planck} Collaboration, P.~Ade et~al., {\it {Planck 2013 results. XVI.
  Cosmological parameters}},  \href{http://xxx.lanl.gov/abs/1303.5076}{{\tt
  arXiv:1303.5076}}.

\bibitem{ATLAS-CONF-2014-009}
{\bf ATLAS} Collaboration, {\it {Updated coupling measurements of the Higgs
  boson with the ATLAS detector using up to 25 fb$^{-1}$ of proton-proton
  collision data}},  {\em ATLAS-CONF-2014-009} (2014).

\bibitem{Chatrchyan:2013zna}
{\bf CMS} Collaboration, S.~Chatrchyan et~al., {\it {Search for the standard
  model Higgs boson produced in association with a W or a Z boson and decaying
  to bottom quarks}},  {\em Phys.Rev.} {\bf D89} (2014) 012003,
  [\href{http://xxx.lanl.gov/abs/1310.3687}{{\tt arXiv:1310.3687}}].

\bibitem{Chatrchyan:2014nva}
{\bf CMS} Collaboration, S.~Chatrchyan et~al., {\it {Evidence for the 125 GeV
  Higgs boson decaying to a pair of $\tau$ leptons}},  {\em JHEP} {\bf 1405}
  (2014) 104, [\href{http://xxx.lanl.gov/abs/1401.5041}{{\tt
  arXiv:1401.5041}}].

\bibitem{CMS:ril}
{\bf CMS} Collaboration, {\it {Updated measurements of the Higgs boson at 125
  GeV in the two photon decay channel}},  {\em CMS-PAS-HIG-13-001} (2013).

\bibitem{Chatrchyan:2013iaa}
{\bf CMS} Collaboration, S.~Chatrchyan et~al., {\it {Measurement of Higgs boson
  production and properties in the WW decay channel with leptonic final
  states}},  {\em JHEP} {\bf 1401} (2014) 096,
  [\href{http://xxx.lanl.gov/abs/1312.1129}{{\tt arXiv:1312.1129}}].

\bibitem{Chatrchyan:2013mxa}
{\bf CMS} Collaboration, S.~Chatrchyan et~al., {\it {Measurement of the
  properties of a Higgs boson in the four-lepton final state}},  {\em
  Phys.Rev.} {\bf D89} (2014) 092007,
  [\href{http://xxx.lanl.gov/abs/1312.5353}{{\tt arXiv:1312.5353}}].

\bibitem{Peskin:1990zt}
M.~E. Peskin and T.~Takeuchi, {\it {A New constraint on a strongly interacting
  Higgs sector}},  {\em Phys.Rev.Lett.} {\bf 65} (1990) 964--967.

\bibitem{Dittmaier:2011ti}
{\bf LHC Higgs Cross Section Working Group} Collaboration, S.~Dittmaier et~al.,
  {\it {Handbook of LHC Higgs Cross Sections: 1. Inclusive Observables}},
  \href{http://xxx.lanl.gov/abs/1101.0593}{{\tt arXiv:1101.0593}}.

\bibitem{Hall:2009bx}
L.~J. Hall, K.~Jedamzik, J.~March-Russell, and S.~M. West, {\it {Freeze-In
  Production of FIMP Dark Matter}},  {\em JHEP} {\bf 1003} (2010) 080,
  [\href{http://xxx.lanl.gov/abs/0911.1120}{{\tt arXiv:0911.1120}}].

\bibitem{Fedderke:2014wda}
M.~A. Fedderke, J.-Y. Chen, E.~W. Kolb, and L.-T. Wang, {\it {The Fermionic
  Dark Matter Higgs Portal: an effective field theory approach}},
  \href{http://xxx.lanl.gov/abs/1404.2283}{{\tt arXiv:1404.2283}}.

\bibitem{Grimus:2008nb}
W.~Grimus, L.~Lavoura, O.~Ogreid, and P.~Osland, {\it {The Oblique parameters
  in multi-Higgs-doublet models}},  {\em Nucl.Phys.} {\bf B801} (2008) 81--96,
  [\href{http://xxx.lanl.gov/abs/0802.4353}{{\tt arXiv:0802.4353}}].

\bibitem{Beringer:1900zz}
{\bf Particle Data Group} Collaboration, J.~Beringer et~al., {\it {Review of
  Particle Physics (RPP)}},  {\em Phys.Rev.} {\bf D86} (2012) 010001.

\bibitem{Degrassi:2012ry}
G.~Degrassi, S.~Di~Vita, J.~Elias-Miro, J.~R. Espinosa, G.~F. Giudice, et~al.,
  {\it {Higgs mass and vacuum stability in the Standard Model at NNLO}},  {\em
  JHEP} {\bf 1208} (2012) 098, [\href{http://xxx.lanl.gov/abs/1205.6497}{{\tt
  arXiv:1205.6497}}].

\bibitem{Antipin:2013sga}
O.~Antipin, M.~Gillioz, J.~Krog, E.~M{\o}lgaard, and F.~Sannino, {\it {Standard
  Model Vacuum Stability and Weyl Consistency Conditions}},  {\em JHEP} {\bf
  1308} (2013) 034, [\href{http://xxx.lanl.gov/abs/1306.3234}{{\tt
  arXiv:1306.3234}}].

\bibitem{Haba:2014sia}
N.~Haba, H.~Ishida, K.~Kaneta, and R.~Takahashi, {\it {Vanishing Higgs
  potential at the Planck scale in singlets extension of the standard model}},
  \href{http://xxx.lanl.gov/abs/1406.0158}{{\tt arXiv:1406.0158}}.

\bibitem{Ade:2014xna}
{\bf BICEP2} Collaboration, P.~Ade et~al., {\it {Detection of B-Mode
  Polarization at Degree Angular Scales by BICEP2}},  {\em Phys.Rev.Lett.} {\bf
  112} (2014) 241101, [\href{http://xxx.lanl.gov/abs/1403.3985}{{\tt
  arXiv:1403.3985}}].

\bibitem{Espinosa:2007qp}
J.~Espinosa, G.~Giudice, and A.~Riotto, {\it {Cosmological implications of the
  Higgs mass measurement}},  {\em JCAP} {\bf 0805} (2008) 002,
  [\href{http://xxx.lanl.gov/abs/0710.2484}{{\tt arXiv:0710.2484}}].

\bibitem{Kobakhidze:2013tn}
A.~Kobakhidze and A.~Spencer-Smith, {\it {Electroweak Vacuum (In)Stability in
  an Inflationary Universe}},  {\em Phys.Lett.} {\bf B722} (2013) 130--134,
  [\href{http://xxx.lanl.gov/abs/1301.2846}{{\tt arXiv:1301.2846}}].

\bibitem{Spencer-Smith:2014woa}
A.~Spencer-Smith, {\it {Higgs Vacuum Stability in a Mass-Dependent
  Renormalisation Scheme}},  \href{http://xxx.lanl.gov/abs/1405.1975}{{\tt
  arXiv:1405.1975}}.

\bibitem{Fairbairn:2014zia}
M.~Fairbairn and R.~Hogan, {\it {Electroweak Vacuum Stability in light of
  BICEP2}},  {\em Phys.Rev.Lett.} {\bf 112} (2014) 201801,
  [\href{http://xxx.lanl.gov/abs/1403.6786}{{\tt arXiv:1403.6786}}].

\bibitem{Barger:2013jfa}
V.~Barger, L.~L. Everett, C.~Jackson, and G.~Shaughnessy, {\it {Higgs-Pair
  Production and Measurement of the Triscalar Coupling at LHC(8,14)}},  {\em
  Phys.Lett.} {\bf B728} (2014) 433--436,
  [\href{http://xxx.lanl.gov/abs/1311.2931}{{\tt arXiv:1311.2931}}].

\end{thebibliography}\endgroup

\end{document}